# Current Perpendicular-to-Plane (CPP) Magnetoresistance (MR).

**by Jack Bass**

**Table of Contents.**

## ABSTRACT

Measurements of Giant Magnetoresistance (GMR) in ferromagnetic/non-magnetic (F/N) multilayers with Current flow Perpendicular to the layer Planes (CPP-geometry) can give better access to the fundamental physics underlying GMR than measurements with the more usual Current flow In the layer Planes (CIP geometry). Because the same measuring current passes through all of the layers, the CPP-MR can often be described by simpler equations that allow separation of effects of scattering within the bulk of the F- and N-metals and at F/N, N1/N2, and F/S (S = superconductor) interfaces. We first describe the parameters that are used to characterize the CPP-MR, the different techniques used to measure these parameters, and the different types of multilayers used to control the two orientations of the magnetizations of adjacent F-layers, anti-parallel (AP) and parallel (P), that permit isolation of the parameters. We then detail what has been learned about the parameters of bulk F-metals, of bulk N-metals, and of F/N, N1/N2, and F/S interfaces. Especially important are the parameters of interfaces and the spin-diffusion lengths in F-metals and F-alloys, about which almost nothing was known in advance. Lastly, we describe work toward CPP-MR devices and studies of magnetothermoelectric effects, before summarizing what has been learned and listing some items not yet understood.

**I. Introduction and Background.**

This chapter covers CPP-MR, the Giant Magnetoresistance (GMR) of magnetic multilayers, composed of alternating layers of ferromagnetic (F) and non-magnetic (N) metals, when the current flows perpendicular to the F/N interfaces (current-perpendicular-to-plane = CPP geometry) [1-7]. Fundamental to GMR is the importance of the spins (magnetic moments) of the conduction electrons. A decade before the discovery of GMR, measurements of Deviations from Matthiessen's Rule in three-component F-based alloys showed [8] that conduction electrons traversing an F-alloy are scattered differently (scattering asymmetry) when their magnetic moments are oriented along or opposite to the moment of the F-alloy. What was new in GMR were: (1) the discovery that such scattering asymmetry could give unexpectedly large MRs in F/N multilayers [9,10], and (2) the discovery that such MRs could be enhanced by similar scattering asymmetries at the F/N interfaces [1-7]. The recognition that electronic transport in magnetic multilayers could depend substantially upon the electron's spin (magnetic moment) gave birth to the name Spintronics, and in 2007 to awarding of a Nobel Prize for the discovery and explanation of GMR.

The intrinsic quantity in the CPP geometry is the specific resistance, AR, the product of the area A through which an assumed uniform CPP current flows and the sample resistance R. Quantitative studies focus on AR for two collinear states: $AR_{AP}$, where the moments of adjacent F-layers are oriented antiparallel (AP) to each other, and $AR_P$, where the moments are oriented parallel (P) to each other. Special interest lies in the difference between the AP and P states, $A\Delta R = AR_{AP} - AR_P$. The CPP-MR is usually defined as CPP-MR = $A\Delta R/AR_P$. $AR_P$ is chosen because it is always measurable, whereas $AR_{AP}$ can be more problematical, as discussed in section IV.

Studies of CPP-MR have focused upon two questions: (1) What is the physics underlying CPP-MR?; and (2) Can CPP-MR be competitive for devices?

Concerning (1), we will argue that the vast majority (maybe all) of published CPP-MR data can be understood in terms of diffuse (as opposed to ballistic) transport, particularized in a one-dimensional model by Valet and Fert (VF) [11]. This model characterizes a multilayer by scattering asymmetries both in the bulk F-layers and at the F/N interfaces. In a properly designed CPP sample, the current density is uniform across the area A. Combining this uniform current density with a collinear orientation of F-layer moments lets the CPP-MR often be analyzed with a one-dimensional model in which layers and interfaces play separate roles. In contrast, the average current in the more usual Current-In-Plane (CIP)-MR (See chapter 1 [12]) flows parallel to the interfaces, and the current density is non-uniform—e.g., for F- and N-layers of comparable thickness, it is larger in the layer with lower resistivity. Separation of contributions of layers and interfaces is usually more difficult. In addition, the characteristic lengths for the CPP-MR (the spin-diffusion lengths $l_{sf}$ [11]) also differ from those for the CIP-MR (mean-free-paths, $\lambda$ [2]), with important consequences for both the magnitudes of the CPP- and CIP-MRs and the equations that describe them. Qualitatively, $\lambda$ is the average distance an electron diffuses between scattering events, whereas $l_{sf}$ is the average distance over which it diffuses between spin-relaxation (spin-memory-loss) events. $l_{sf}$ is usually several times longer than $\lambda$, because spin-relaxation events are typically only a small fraction of scattering events. We will show below that these differences can lead to equations for the CPP-MR that allow relatively direct separations of the bulk and interface contributions to GMR. Of the important parameters, the least known before CPP-MR studies were the parameters of interfaces, and the spin-diffusion lengths in F-metals and F-alloys. We emphasize quantitative results, including examples where parameters derived from CPP-MR measurements agree well with equivalent ones determined by completely different techniques, and/or agree well with no-free-parameter calculations.

Concerning (2), it was shown early on that the CPP-MR of a simple $[F/N]_n$ multilayer ($n$ is the number of bilayer repeats) could be several times larger than the CIP-MR, both at 4.2K (Fig. 1 [1]) and up to room temperature (Fig. 2 [4,13,14]). Such ratios are consistent with calculations [15]. However, the CPP-MR has two disadvantages for devices. (a) The resistance R of a standard CIP-MR multilayer measured in the CPP geometry is tiny, $R \sim 10^{-8}\ \Omega$ for CPP length ~ 1 μm and A = 1 $mm^2$. To give large enough R for devices, standard metallic CPP multilayers must have areas < $10^{-2}$ $(\mu m)^2$, requiring nanolithography. (b) Devices such as read heads require the CPP multilayer to be short to match the bit size. In CPP-MR, the lead resistances, and layers used to 'pin' F-layer magnetizations, are in series with the active CPP-MR components. With standard F-metals and alloys these series components limit the CPP-MR. Recently, fabrication of nanopillars with new F-compounds and combinations of materials have produced CPP-MRs more competitive for sensors, as will be discussed in section IX.

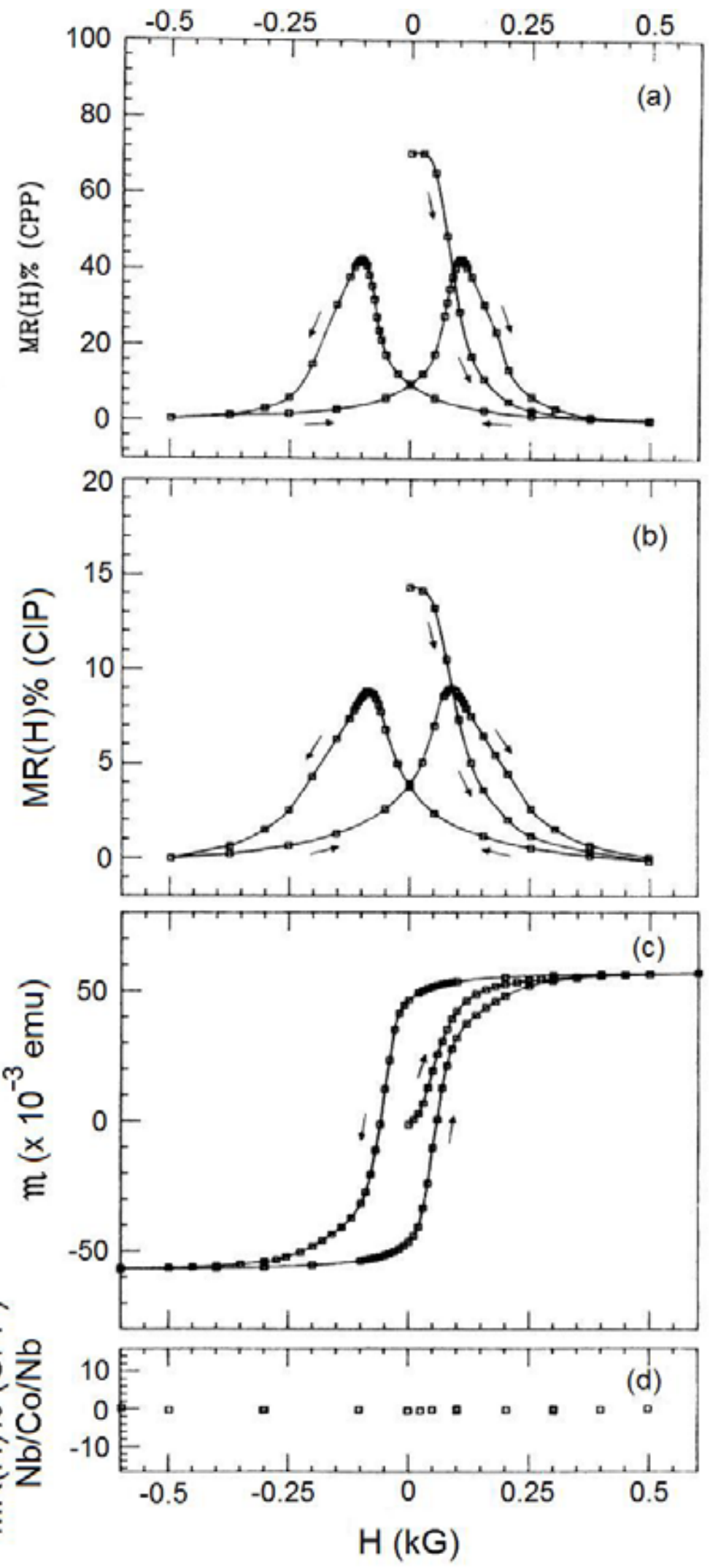

**Fig. 1**. (a) CPP-MR%, (b) CIP-MR%, and (c) magnetization *M*, vs magnetic field H for sputtered $[Co(6)/Ag(6)]_{60}$ multilayers at 4.2K. (d) CPP-MR for a Nb/Co(200)/Nb film. Multilayer specific resistances are defined for three states: The largest value, $AR_o$, is for the as-prepared state at H = 0, the intermediate value, $AR_{Pk}$, is at the peak following saturation, and the smallest value, $AR_s$ is at saturation. In the text we will use and justify the approximations: $AR_o = AR_{AP}$ and $AR_s = AR_P$. From W.P.Pratt Jr. et al., [1].

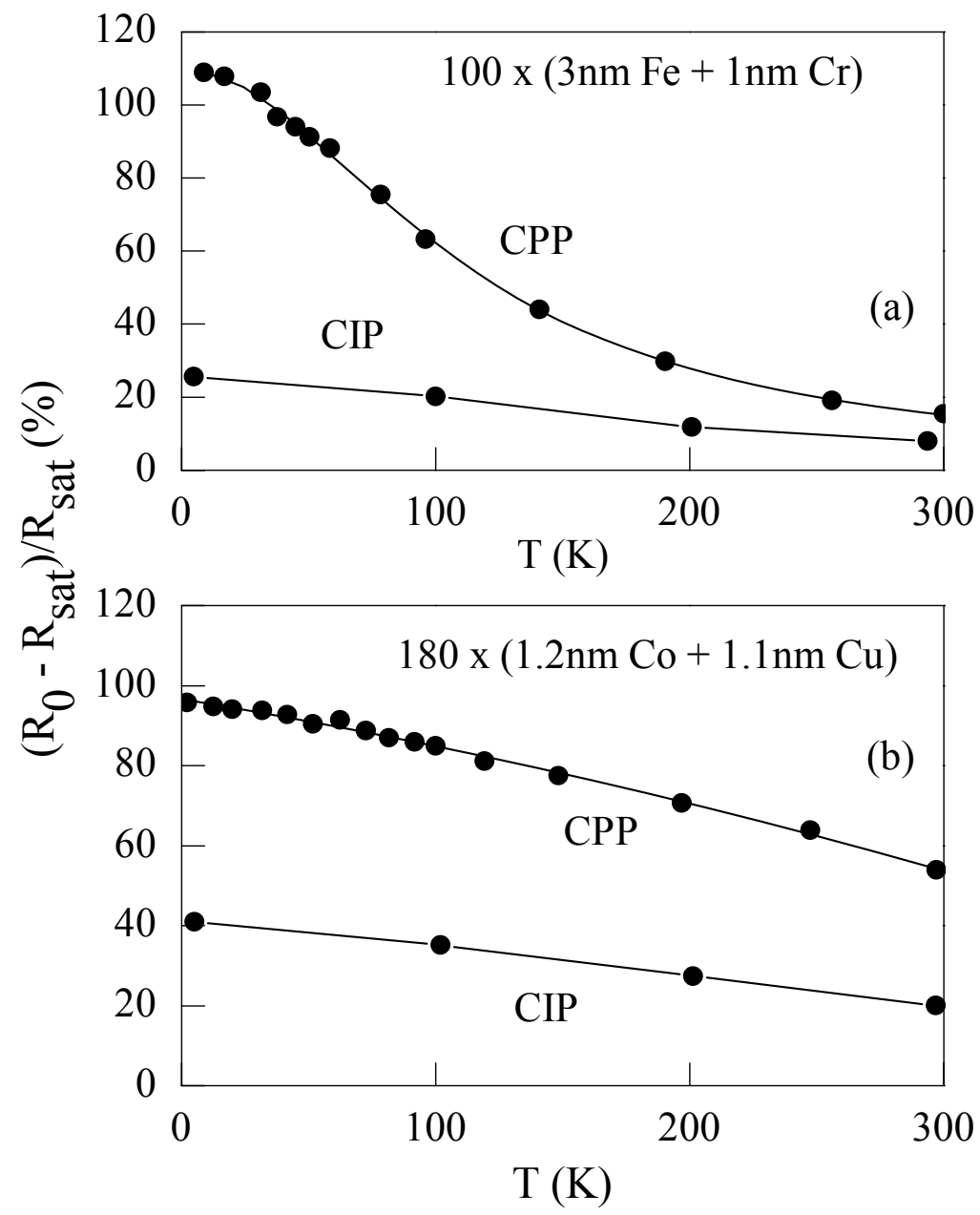

**Fig. 2.** $\Delta R/R_P$ vs temperature T, for nanopillar multilayers of : (a) FeCr, and (b) Co/Cu. From J.Bass and W.P. Pratt Jr., [4]; After M.A.M. .Gijs et al. [13,14].

chapter is organized as follows. Section II presents the parameters that characterize the CPP-MR. Section III describes the three different techniques that have been used to measure the CPP-MR, along with their advantages and disadvantages. Section IV explains different ways to obtain the anti-parallel (AP) order of magnetic moments that generally gives the largest AR or MR. Section V briefly describes the theory of Valet and Fert (VF) [11] that is most often used to analyze CPP-MR data, and explains what is involved in realistic calculations of CPP-MR parameters. For more extensive discussions of CPP-MR theories and their limitations, see [2,3,6]. In the limit of no spin-relaxation, VF theory reduces to a simple two-current series-resistor (2CSR) model [11,16]. Examples of especially useful 2CSR equations, or VF equations in appropriate limits, are given along with experimental data in sections VI-VIII. Section VI describes a series of tests that were made to test the 2CSR and VF models. Section VII covers the bulk parameters derived from CPP-MR measurements. Section VIII covers the interfacial parameters. Section IX describes progress toward CPP-MR devices. Section X briefly covers magneto-thermoelectric effects and thermal conductance. Section XI summarizes what we see as the most important CPP-MR results obtained so far, and notes some topics that are not yet understood..

Space limitations preclude our describing all of the limitations on the many measurements, assumptions, and analyses of CPP-MR data in the literature, of which we can cover only some. So we just warn that published claims and parameters must be viewed with caution, and list here a set of questions worth asking. Are there enough different data sets to determine the required unknowns? Typically a single data set (e.g., $A\Delta R$ vs F-layer thickness $t_F$ , or vs the number *n* of F/N bilayers) can reliably determine only 2 or perhaps 3 unknowns. To determine more, and show that the resulting values are not functions of the variables (e.g., that AR doesn't vary with *n*), requires more than one data set, or independent measurements to fix other parameters. Does the technique used produce good AP and P states? Is current flow through the multilayer uniform? If a 2CSR model is used, is spin-relaxation negligible, including in the contacts? If a VF model is used, are the parameters of the contacts known and properly included, and are all fixed parameters measured in the same laboratory? Using parameters derived by other groups with different sample preparation systems is usually unreliable, especially for nominally 'pure' metals.

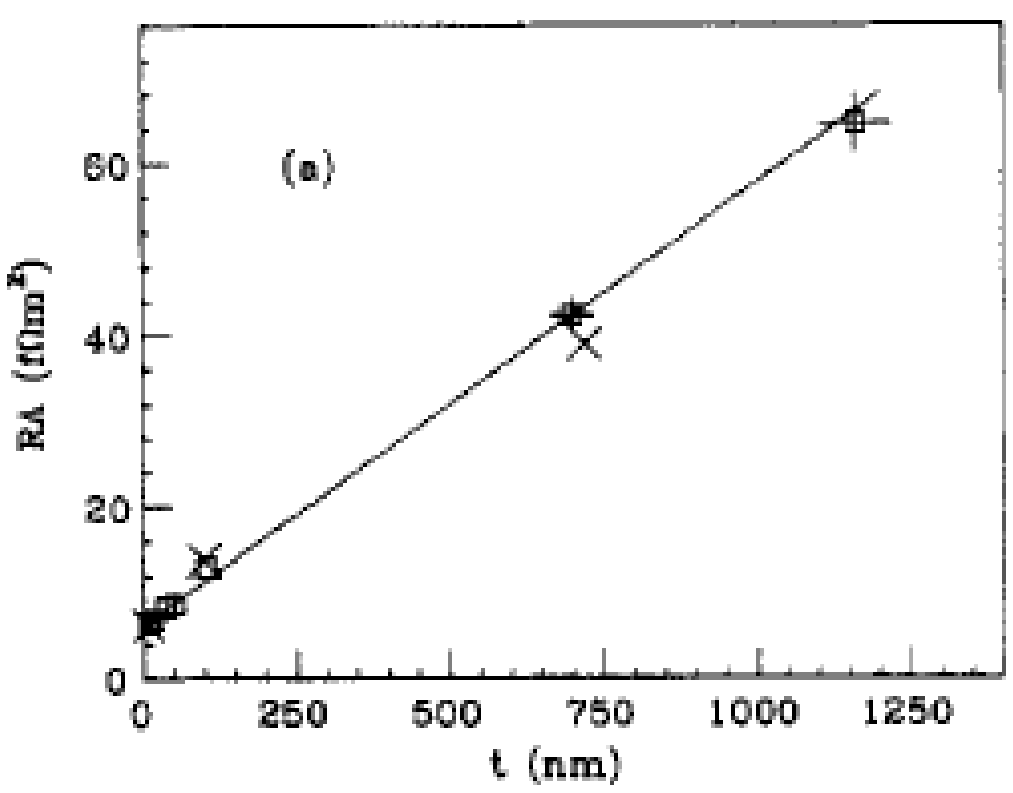

**Fig. 3.** AR vs Co layer thickness t at 4.2K for CPP-current flow into sputtered Nb/Co(t)/Nb trilayers with superconducting Nb contacts. The x's are for samples with 10 nm of Ag at the interfaces between the Nb and Co. The ordinate intercept gives $2AR_{Co/Nb}$. The slope gives $\rho_{Co}$. From C.Fierz et al. [17].

## II. CPP-MR Parameters.

The physics underlying GMR for a simple F/N/F trilayer is explained in Chapter 1 [12]. We summarize here only those features that are essential for the CPP-MR. Because the spin of an electron is 1/2, the electron's magnetic moment can be quantized into two states along any chosen axis, such as the axis of an applied magnetic field H. We call the two states 'up' and 'down'. A conduction electron suffers different amounts of scattering when its moment is along or opposite to the moment of an F-layer that it is traversing. For diffuse transport, this scattering within F is characterized by parameters $\rho_F^{\uparrow}$ and $\rho_F^{\downarrow}$, where $\uparrow$ means that the electron moment points along the F-layer moment and $\downarrow$ means that it points opposite to the F-layer moment. Values of these two parameters for F-based binary alloys were already estimated years earlier from Deviations from Matthiessen's Rule (DMR) studies of asymmetric scattering in F-based ternary alloys [8]. Usually $\rho_F^{\downarrow} > \rho_F^{\uparrow}$. The CPP equations, especially that for $AR_{AP}$, can be simplified by using an alternative pair of parameters (first defined in [11,16]), as shown in section VI.B below. These parameters are the dimensionless scattering asymmetry, $\beta_F = (\rho_F^{\downarrow} - \rho_F^{\uparrow})/(\rho_F^{\downarrow} + \rho_F^{\uparrow})$--bounded by $-1 \leq \beta_F \leq 1$, and the enhanced resistivity, $\rho_F^{*} = (\rho_F^{\downarrow} + \rho_F^{\uparrow})/4 = \rho_F/(1-\beta_F^2)$. Here $\rho_F$ is the resistivity of the F-metal as measured independently, either by measuring the slope of a plot of the CPP AR vs $t_F$ for F-layers of variable thickness $t_F$ (Fig. 3) [17], and/or in the CIP geometry using the Van der Pauw method on films deposited in the same way as the multilayers and thick enough to minimize effects of surface scattering. For Co and Ni, Fierz et al. [17] found that the values of $\rho_F$ measured in these two different ways overlapped to within mutual uncertainties, providing some confidence in the latter technique.

Similarly, scattering at an F/N interface is characterized by the parameters $AR_{F/N}^{\downarrow}$ and $AR_{F/N}^{\uparrow}$. For CPP-AR analysis, these can be combined to give the alternative dimensionless interface scattering asymmetry $\gamma_{F/N} = (AR_{F/N}^{\downarrow} - AR_{F/N}^{\uparrow})/(AR_{F/N}^{\downarrow} + AR_{F/N}^{\uparrow})$—also bounded by -1 and 1, and the enhanced interface specific resistance, $AR_{F/N}^{*} = (AR_{F/N}^{\downarrow} + AR_{F/N}^{\downarrow})/4$.

Lastly, scattering within the N-layer is characterized by just $\rho_N$, since such scattering should be independent of the direction of the electron's moment. As with $\rho_F$, $\rho_N$ can be measured separately either in the CPP-geometry (but, when superconducting contacts are used, thin F-layers must be included as bookends on N to avoid a superconducting proximity effect on the N-metal), or in CIP with the Van der Pauw technique. For N-metals with low resistivities (e.g., Cu and Ag), layer thicknesses in multilayers can sometimes be shorter than the mean-free-path, raising the possibility of ballistic transport. The available evidence is that no significant change in CPP-MR occurs as the layer thicknesses are reduced below this boundary. Presumably incoherent, diffusive scattering dominates the CPP-MR for two coupled reasons: (a) the contribution to AR from such thin, pure N-layers is often too small to matter, and (b) coherent effects are eliminated by diffuse scattering within disordered and rough interfaces.

If, as electrons propagate through a multilayer, their moments don't flip, then currents of 'up' and 'down' electrons propagate independently, giving a two-current (2C) model, where the conductances for 'up' and 'down' electrons simply add [15]. If transport within the multilayer is also diffuse, the total specific resistance for each current is simply the sum of the contributions from the local resistivity ( $\rho_F^{\uparrow}$ , $\rho_F^{\downarrow}$ , or $\rho_N$,) for a given layer times the layer thicknesses ($t_F$ or $t_N$), and the contributions from the interface specific resistances ( $AR_{F/N}^{\downarrow}$ or $AR_{F/N}^{\uparrow}$ ). These sums give the Series-Resistor (SR) model. Combining the 2C and SR models gives the 2CSR model, examples of which will be given in sections VI-VIII.

In sections VI.E,F, VII, and VIII we will also examine what happens when the moments of the electrons flip as the electrons traverse the layers and interfaces of a multilayer. At low temperatures, scattering is just from impurities, which produce large angle scattering. On average, such scattering randomizes the final crystal momentum. If so, when spin-relaxation also occurs, due to spin-orbit scattering from impurities without local moments (most impurities), or to spin-spin scattering from impurities with local moments (e.g., Mn [18,19]), such flipping does not transfer crystal momentum to the other spin-channel and, thus, does not mix currents [11,19]. We call spin-flipping that does not mix currents 'spin-relaxation'. When spin-relaxation is present, the 2CSR model must be generalized to the VF model, which is still a two-current (2C) model, but no longer a series-resistor (SR) one. To describe such relaxation requires the following additional parameters, the spin-diffusion lengths, $l_{sf}^F$ and $l_{sf}^N$ , within the F- and N-metals [11], and the spin-relaxation parameters, $\delta_{N1/N2}$, $\delta_{F/N}$, and $\delta_{F1/F2}$, at N1/N2, F/N, or F1/F2 interfaces [20,21,22]. Crudely, $l_{sf}^F$ and $l_{sf}^N$ are the lengths over which conduction electron spins relax within the F- and N-metals (the lengths over which the spin-accumulation varies), and δ specifies the probability $P = (1 - \exp(-\delta))$ that a conduction electron's spin flips (relaxes) as the electron crosses a given interface. Usually, spin-relaxation leads to reduction of AΔR. In the simplest cases, AΔR can decrease as $\exp(-t/l_{sf})$ or as $\exp(-\delta)$. In contrast, at higher temperatures, electron-electron, electron-magnon, and electron-phonon scattering can lead to spin-flipping with transfer of momentum to the other spin-channel, which we call spin-mixing. An additional spin-mixing parameter is then needed to determine how the spin-currents mix, again usually reducing AΔR. So far, information about spin-mixing from both CPP-MR calculations and measurements is modest [23,19].

Before turning to techniques and data, we briefly consider when bulk and interface parameters are intrinsic or extrinsic.

We start with the bulk parameters. The values of $\beta_F$ for dilute F-alloys estimated from measurements of Deviations from Matthiessen's Rule [8] vary substantially for different impurities, from $\beta_F \sim -0.8$ for V in Fe to $\beta_F \sim +0.85$ for Fe in Co. With such a wide variation in $\beta_F$, it seems clear that $\beta_F$, $l_{sf}^F$ (and $l_{sf}^N$ ) are well defined only for F-alloys (or N-alloys) in which a single, known impurity dominates the scattering. We'll see in Section VII that CPP-MR values of $\beta_F$ for several F-alloys agree reasonably well with DMR values. In contrast, the scattering from the expected impurities in F-metal or N-metal targets with specified purities of 99.9% or better, is much smaller than needed to explain the observed residual resistivities, $\rho_F$ or $\rho_N$, of films deposited from such targets. Thus the dominant impurities or defects in thin films of nominally pure metals such as Co, Fe, Ni, Cu, and Ag are unknown, and values of $\beta_F$, $l_{sf}^F$ , or $l_{sf}^N$ derived for deposited layers of one of these 'pure' metals can be only an approximation for layers of that metal, deposited by that group, with reasonably stable values of $\rho_F$ or $\rho_N$. The rough agreement that we will see below for values of $\beta_{Co}$ derived by different groups, with a wide range of values of $\rho_{Co}$, is thus rather a surprise.

In contrast, the F/N interface parameters— $\gamma_{F/N}$, $2AR_{F/N}^{*}$ , and $\delta_{F/N}$ (and similar parameters for N1/N2 or F1/F2 interfaces)—might be determined solely by the properties of the two metals, if the parameters are not sensitive to the detailed structure of the interface (e.g., whether the interface is a perfect plane, or consists of a finite thickness of an interfacial alloy--often 3-4 monolayers (ML) thick [24]), or contains physical surface roughness). We will see in section VIII that the scale of values for 2AR is '$f\Omega m^2$' (or $m\Omega(\mu m)^2$)—that is, values of 2AR vary from $\sim 0.1\ f\Omega m^2$ to $\sim 10\ f\Omega m^2$, and that calculated values of 2AR for some metal pairs are not highly sensitive to intermixing. For lattice matched pairs (same crystal structure and closely the same lattice parameters), we'll see that no-free-parameter calculations of 2AR agree rather well with experimental values.

**III. Measuring Techniques.** Three different F/N sample geometries (listed in order of first publications) have been used to measure the CPP-MR: (1) Short-wide multilayers, sandwiched between crossed superconducting strips [1,25,26]; (2) Multilayer pillars with more closely comparable width and length [13]; and (3) Multilayer nanowires with lengths much longer than widths [27-30]. By itself, the geometry of only case (3) guarantees a uniform current density through the wire. In case (1), two superconducting strips are needed to give equipotential surfaces, even when current is flowing, to make the current density uniform [26], in analogy with why the electric field is uniform in a short-wide capacitor of two metal strips sandwiching an insulator. In case (2), the current density is generally not strictly uniform, because the contacts are not strictly equipotentials. But, with care, the deviations from uniformity can often be controlled or corrected for.

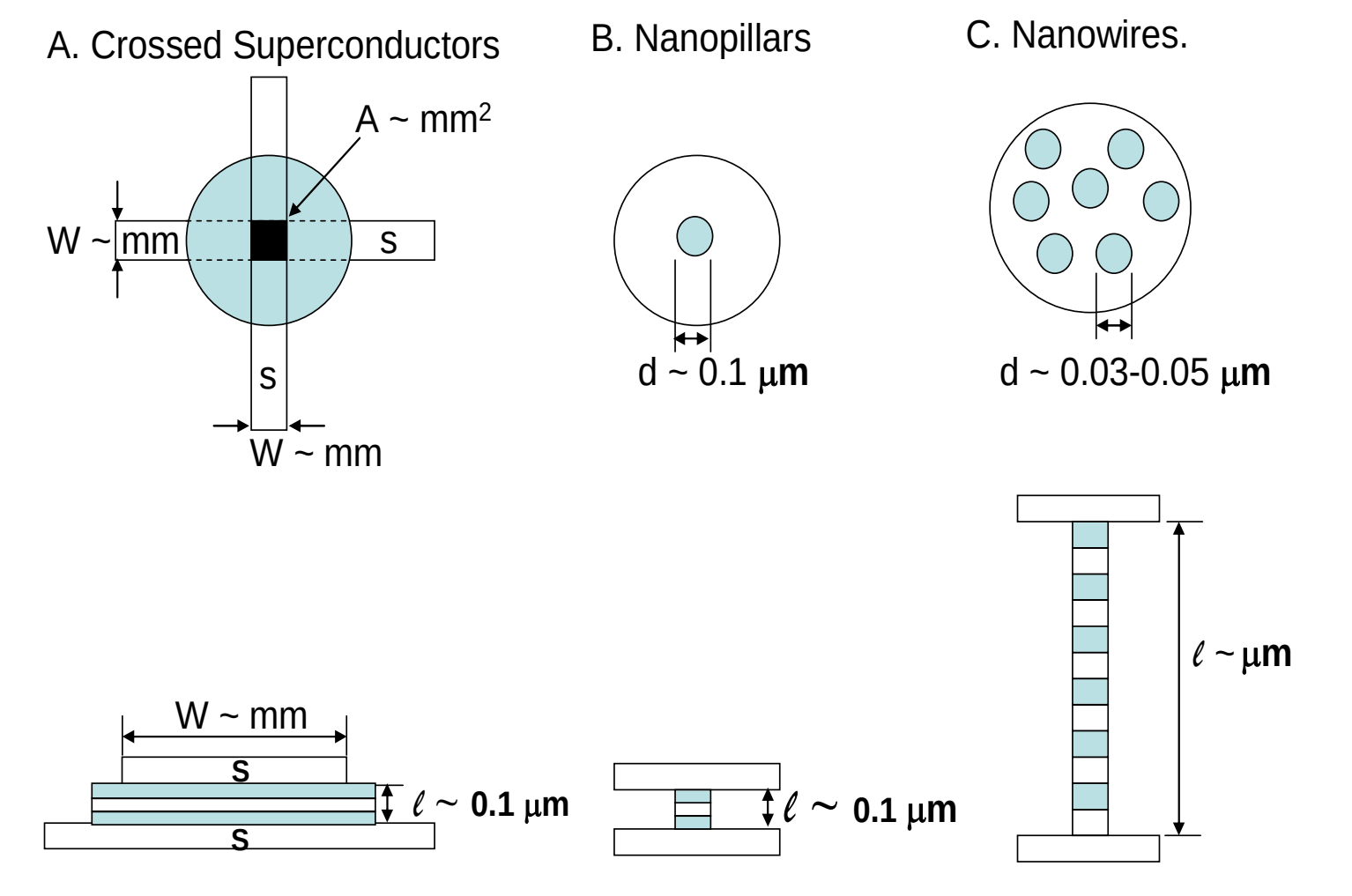

**Fig. 4**. CPP-MR Measuring Geometries: Top View Above. Side View Below. Not to scale.

In all three techniques, non-epitaxial sputtering or electron-beam evaporation standardly give closest packed layer planes (i.e., (111) planes for fcc or (011) planes for bcc). The separation between (111) planes in fcc is ~ 0.2 nm.

With this background we now discuss each of the three techniques in more detail.

(1) The first technique [1] involves sandwiching a thin ( ≤ 1 micron) multiayer of interest between mm-wide crossed superconducting strips of Nb (Fig. 4A). This technique was first tried by Schuller and Schroeder [31] at Argonne Natl. Laboratory. But the need to open the sputtering system to air between deposition of the Nb strips and the multilayer caused uncontrolled interfacial oxidation. Returning home to Michigan State University (MSU), Schroeder and his colleague Pratt designed an ultra-high vacuum compatible sputtering system with in-situ mask changing [25,26] that allowed masks to be changed in minutes without breaking vacuum. Advantages of the technique include the following. (A) As noted above, this short, wide sample geometry gives a uniform current density. (B) Multilayers can be deposited with arbitrary combinations of F-, N-, and anti-ferromagnetic (AF) metals, allowing control of both AP and P states and studies of many different F and N combinations. (C) Zero resistance of the superconducting Nb strips simplifies the contact resistance, which is just $AR_{S/F}$ the interface resistance between the S and F metals. This simplicity is convenient for data analysis, since contact resistances can be important in CPP-MR. (D) $2AR_{S/F}$ can be measured independently, as the ordinate intercept of a plot of AR vs $t_F$ for sandwiches of F-metal thickness $t_F$ between the S leads (Fig. 3) [17]. The slope of such a plot also gives $\rho_F$. (E) Measurements at 4.2K avoid contributions from phonons and magnons, making it easier to compare data with calculations. Because of these advantages, most of the quantitative analyses that we discuss below were made with this technique. The main disadvantages of the technique are the following. (A) The need for a high sensitivity, high precision bridge system [26,32] to measure the resulting very small resistances (~ 10 nΩ). (2) Its limitation to cryogenic temperatures (T ~ 4.2K with Nb), although we'll see that interfacial parameters are most likely not very temperature sensitive. Two related techniques have been published: (a) Still using a precision bridge, Slater et al. [33] used superconducting contacts to pillars as small as micron diameter. (b) To allow measurements with a commercial digital voltmeter, Cyrille al. [34] sputtered, in series, one-hundred, ~ 30 μm diameter, multilayers with Nb contacts between them.

(2) The second technique [13,14] involves sputtered or evaporated multilayers, shaped into micro- or nano-pillars by optical or electron-beam lithography (Fig. 4B). The main advantages of this technique are the following. (A) Multilayers can be deposited with arbitrary combinations of F-, N-, and AF-metals; (B) Measurements can be extended from 4.2K to above room temperature; (C) Resistances are large enough to measure with standard digital voltmeters. The main disadvantages are: (A) The need for complex optical and/or nanolithography to produce good samples. (B) Difficulty in achieving near equipotentials across the top and bottom contacts, to assure uniform current density through the pillar. The first measurements with micron$^2$ areas had clear problems with non-uniform currents [35]. (C) Contact resistances that can be comparable to the multilayer resistances, and difficult to determine for inclusion in proper VF analyses. Contact resistance and non-uniform current problems are examined in supplementary note #1.

(3) The third technique involves electrodepositing 40-100 nm diameter (d) nanowires into either polymers, with pores etched after their axes are defined by ion-bombardment [27-30]), or Al-oxide with pores made by etching [36](Fig. 4C). The advantages of this technique are the following. (A) The current density is uniform. (B)

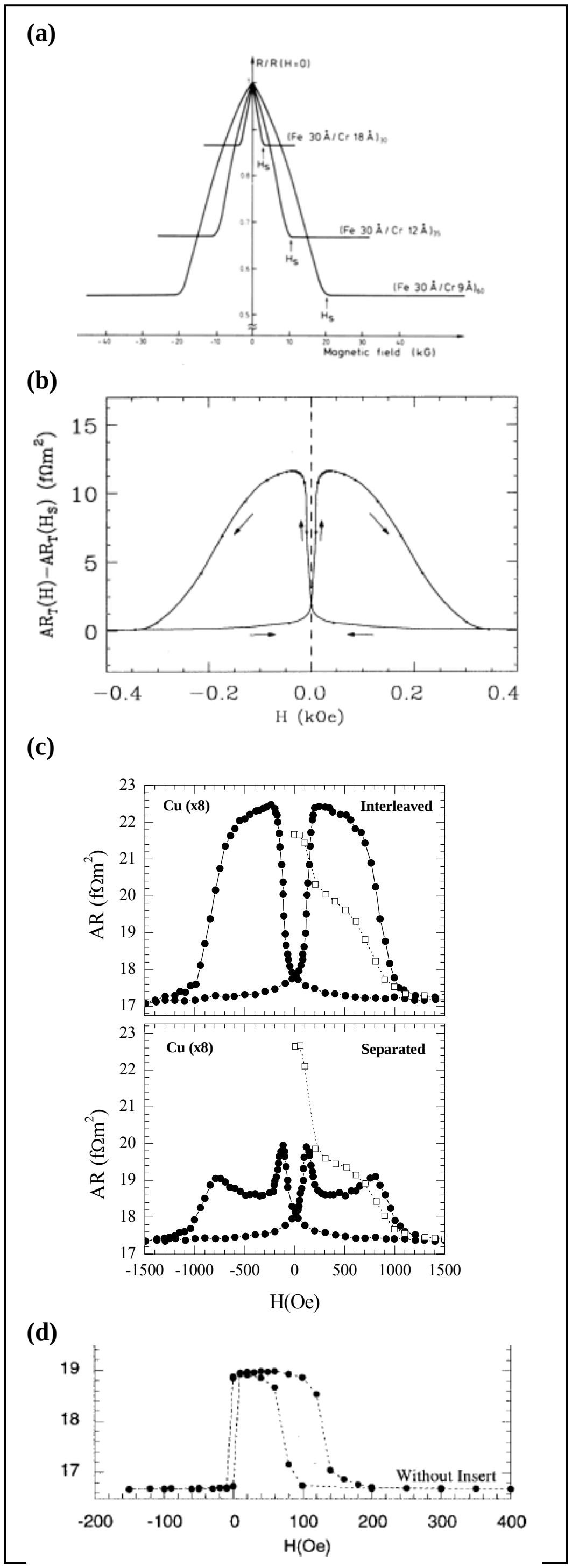

**Fig. 5**. Hysteresis curves for multilayers with well-defined AP states. (a) Fe/Cr with antiferromagnetic coupling. (b) $[Co(3)/Cu(20)/Py(8)/Cu(20)]_8$ hybrid spin-valve. (c) Interleaved ($[Co(1)/Cu(20)/Co(6)/Cu(20)]_6$ vs Separated $[Co(1)/Cu(20)]_6[Co(6)/Cu(20)]_6$ hybrid spin-valves; (d) Py-based exchange-biased spin-valve (EBSV). From: (a) M.N. Baibich et al. [9]; (b) Q. Yang et al., [50]; (c) K. Eid et al. [53], (d) W. Park et al. [20]

Resistances are large enough to measure with standard digital voltmeters. (C) Measurements can be extended from 4.2K to above room temperature. (D) The long thin geometry allows significant temperature gradients to be established. So this geometry has been used to pioneer studies of thermoelectric GMR [37,38]. The disadvantages are the following. (A) Most published work involves deposition of the F and N-metals from a single bath, limiting the wires to just simple F/N multilayers with a limited number of F and N pairs [5]. The N-metal has usually been Cu, and most data involve either Co/Cu or Py/Cu (Py = Permalloy = $Ni_{1-x}Fe_x$ with x ~ 0.2) [5]. Usually, the metal deposited at lower voltage—e.g. Cu, contaminates the metal deposited at the higher voltage—e.g., Co. (B) Most studies involve contacts to multiple wires of unknown number. Then, measurements are reported only of MR. Few studies have been reported with single wires [e.g., 39,40]. (C) Most studies have involved equal thickness F-layers and equal thickness N-layers, making it difficult to achieve fully AP states. A complication is that the magnetic orientations of the F-layers change as the layer thickness $t_F$ increases through the wire diameter d. For $t_F < d$, shape anisotropy drives the F-layer moment in-plane, and the dipolar interaction between adjacent F-layers is antiferromagnetic. For $t_F > d$, in contrast, shape anisotropy drives the moment along the wire axis, and the dipolar interaction is ferromagnetic. (D) A few studies have been made with multiple baths. In the first, Co/Cu multilayers prepared in two-baths [41] gave lower MRs than ones prepared in a single bath, a difference attributed mostly to lesser contribution from Co/Cu interfaces. More recently, multiple baths, and inclusion of an antiferromagnet (AF) to give pinning, have given exchange-biased spin-valves (EBSVs) [42]. While some of the results look reasonable, others do not.

A few measurements have been made on samples with grooved surfaces, giving current at an angle to the plane (CAP) [see, e.g., 43-49]. The main contributions so far to CPP-MR are: (a) evidence that $\beta_F$, $\gamma_{F/N}$ and $AR_{F/N}$ are only modestly sensitive to temperature (≤ 10-20% from 4.2K to 300K) [47], a conclusion supported by a subsequent nanowire study [23], and (b) early data on magnetothermopower in CAP and pseudo-CPP geometries [48,49].

**IV. Determining AΔR or MR: Control of AP and P states.** The P-state is usually achieved by just increasing the magnetic field H to above the saturation field of the F-layer with the largest saturation field. The moments of all of the F-layers should then point in the direction of H, giving $AR_P$.

Obtaining $AR_{AP}$ requires more care. Several methods have been used. (1) GMR was discovered in Fe/Cr multilayers [9,10] where the Cr thickness was chosen to give antiferromagnetic (AF) coupling between neighboring Fe layers Fig. 5a [9](See Chapter 1). This method has the disadvantages of allowing samples with only one or two thicknesses of the Cr layers, and requiring large H to reorient to the P-state. (2) In the first CPP-MR studies, on $[Co/Ag]_n$ multilayers, with Ag layers thick enough to make exchange coupling weak, the initial values of AR in as-prepared samples, not yet subjected to H, were found to be the largest achievable (Fig. 1a) [1], typically much larger than the values near the coercive field. These initial values were taken as the best available approximations to $AR_{AP}$. They were subsequently validated in two ways. In the earliest, they were able to predict, correctly, with no adjustability, values of AΔR for $[Co/Cu/Py/Cu]_n$ multilayers, which give well defined AP states because of the very different coercive fields of the Co and Py layers [Fig. 5b][50,51]. Later, a combination of polarized neutron scattering and Scanning Electron Microscopy with Polarization Analysis (SEMPA) showed that the initial state could indeed approximate the AP state in Co/Cu multilayers [52], in that the typically micron sized or larger domains in a given Co layer had moments oriented approximately opposite (to ≥ 90%) to the coupled domains in the layers just above and just below it. This correlation was attributed to antiferromagnetic ordering produced during growth of a given layer by the fringing fields of the domains in the layer preceding it [52]. (3) $[F1/N/F2/N]_n$ multilayers with F1 and F2 having different coercive fields, $H_c$. F1 and F2 can be different metals—e.g. Co and Py as above, or different thicknesses of the same F-metal (interleaved sample in Fig. 5c) [53]. We call such samples 'hybrid' spin-valves (SVs) (4) Exchange-bias with an AF—F/N/F/AF, to 'pin' the moment of the F layer adjacent to the AF layer so that it reverses at a much higher field than does the other 'free' F-layer. This procedure gives an exchange-biased spin-valve (EBSV) (Fig. 5d) [20]. The exchange bias is produced by heating the multilayer to above the blocking temperature of the AF, and then cooling in the presence of a field H [4]. (5) In long nanowires, with diameters larger than F-layer thicknesses, an AP state can be obtained by alternating N-layer thicknesses between a short value that lets the moments of the two bounding F-layers orient antiparallel (AP) due to their dipolar coupling, and a long value that magnetically separates such bonded pairs (see, e.g. Fig. 19 in ref. [5]). (6) As shown in Fig. 4b, in a nanopillar with only two thin F-layers separated by a not-too-thick N-layer, dipolar coupling between the two F-layers will orient their magnetizations AP at H = 0. Methods (3) – (6) have the advantage of allowing controlled AP-states with combinations of a wide variety of F and N metals.

## V. Theory Overview.

Wide ranging reviews of CPP-MR theory are given in Levy [2] and Gijs and Bauer [3]; and a more focused one in Tsymbal and Pettifor [6]. Topics covered include comparisons between Boltzmann Transport theory, Kubo theory, and Landauer formalism, as well as differences between ballistic and diffuse scattering. In this review, we focus upon the model used to analyze nearly all CPP-MR data, the Valet-Fert (VF) model [11]. This model assumes diffuse transport based upon the Boltzmann equation, and reduces in the limit of no-spin-relaxation to a two-current series-resistor (2CSR) model.

The first specific model of CPP-MR was gjven by Zhang and Levy [15]. Neglecting spin-flip scattering, they argued that 'each of the two spin-directions (since the electron spin is ½) 'contributes independently', giving a total conductance that is just the sum of their separate conductances (i.e., a two-current (2C) model). They then showed that the CPP resistance for each spin-channel is 'self-averaging'; that is, it is just the sum of the resistance contributions from the layers and interfaces (i.e., a series-resistor (SR) model). Together, these two results predict a 2CSR model that was the model used to interpret much early CPP-MR data. This model contains no lengths beyond just the layer thicknesses.

As noted in section II, neglecting contacts, the 2CSR model for a simple [F/N] multilayer has only five parameters, $\rho_N$, $\beta_F$, $\rho_F^* = \rho_F/(1-\beta_F^2)$, $\gamma_{F/N}$, and $AR^*_{F/N}$, of which $\rho_N$ and $\rho_F$ can be measured independently, leaving only three unknowns. However, contacts usually require at least one more parameter, an example of which will be given in section VI.

### (A) Valet-Fert Theory of Diffuse Scattering with Spin-Relaxation.

Soon after Zhang and Levy [15], it was recognized that spin-relaxation need not be negligible in real F/N multilayers [11,54]. The Valet-Fert (VF) model [11] used to fit most experimental data also starts from the Boltzmann equation. VF assumed the same, single band, spherical Fermi surface for both the F and N-metals, and their analysis is formally valid only in the limit $l_{sf} >> \lambda$. However, the form of their equations is expected to apply more generally [55], and comparing VF with numerical solutions of the Boltzmann equation led Penn and Stiles [56] to conclude that the VF equations remain (closely) valid even when $l_{sf}$ is only comparable to λ. When $l_{sf} >> \lambda$, VF first showed that the Boltzmann Equation reduces to a macroscopic model in which current densities are related to electrochemical potentials [57,58]. The characteristic lengths in the model are $l_{sf}^N$ and $l_{sf}^F$. They then derived a 'spin-diffusion type' equation for the spin-accumulation, which led to general solutions (with the CPP-MR parameters listed in section II) for the chemical potentials, electric fields, and currents within the layers of the multilayer. Finally, they gave equations for these quantities within the F- and N-layers and specified how to match boundary conditions at the F/N interfaces, including the localized spin-dependent interface resistances $AR^{\downarrow}_{F/N}$ and $AR^{\downarrow}_{F/N}$ defined above. They also provided examples of solutions for some simple cases. The first solutions were for a single F1/F2 interface, and for a simple $[F/N]_n$ multilayer, both with zero interface resistances. The last solutions were for a general periodic multilayer including spin-dependent interface resistances. Because of the complexities associated with different 'contacts', they didn't give any solutions for samples with realistic contacts, leaving it to the experimenter to apply the Valet-Fert (VF) equations within the F- and N-layers, and the VF matching of boundary conditions, to real data. In samples with a variety of F-, N-, and possibly also AF-layers, plus 'contacts', applications of the VF model will usually require complex numerical fits [20]. We'll give below some examples where the VF analysis, including superconducting contacts, reduces to relatively simple equations.

The VF analysis did not include the parameter δ that describes spin-relaxation at a metallic interface. Park et al. [20] first introduced δ into a VF analysis by treating each interface as a slab of finite thickness $t_I$, with resistivity $\rho_I$, spin-diffusion length $l_I$, and $\delta = t_I/l_I$. These slabs were incorporated into the VF analysis as additional 'layers'. Results of such analyses will be given in section VIII.D.

**(B) Realistic Calculations.**

To calculate the VF parameters for F and N metals, requires use of real electronic structures (Fermi surfaces). The best agreement so far between measured VF parameters and no-free-parameter calculations occurs for calculations of twice the interface specific resistance, $2AR_{N1/N2}$ or $2AR^{*}_{F/N}$, for lattice matched metal pairs--i.e., pairs with the same crystal structure (fcc or bcc) and the same lattice parameter to within ~ 1%. Lattice matching lets a common crystal lattice be used for the two metals forming the interface. Consistent with the VF assumption that CPP electron transport in multilayers is diffuse, ref. [59] showed that assuming ballistic bulk transport gave results in strong disagreement with experiment for Co/Cu interfaces. In contrast, assuming diffuse bulk scattering gave good agreement. Presumably, interfacial disorder precludes the coherent scattering between neighboring interfaces that would be expected for ballistic transport. Calculating $2AR_{N1/N2}$ or $2AR^{*}_{F/N}$ for a lattice matched pair requires two steps. The first involves determining the electronic structure for each metal self-consistently within the local spin density approximation. The second involves calculating the interface specific resistance using an appropriate equation for a single interface based upon Landauer theory, corrected for the Sharvin resistance [59,60]. The calculated results given in section VIII below were obtained for two kinds of interfaces. The first is a perfect interface with specular scattering. Here, transport across the interface requires conservation of the component of the wave-vector **k** parallel to the interface. The second is a 50%-50% random mixture of atoms 2 monolayers (ML) thick. Now transport across the interface involves both a specular component and a diffuse component where **k** parallel is not conserved. Early calculations used a basis set with spd and linear muffin-tin orbitals (LMTO) [61-63]. Later calculations used spdf and MTO orbitals without linearization [64]. The spdf and MTO results will be given in section VIII.

## VI. Tests of the 2CSR and VF theories.

### (A) Introduction.

An early task in CPP-MR studies was to test whether real data can be consistent with the simple 2CSR and VF models. In this section, we describe the results of some such tests, which were made using the crossed superconductor geometry.

The first CPP-MR study showed that the CPP-MR for $[Co/Ag]_n$ multilayers is typically several times larger than the CIP-MR [1], as illustrated in Fig. 1 for a $[Co(6nm)/Ag(6nm)]_{60}$ multilayer. The first detailed analysis of CPP-MR data, on $[Co/Ag]_n$ multilayers, assumed a one-current series resistance model [65]. Soon afterward, extension of measurements to $[Co/AgSn]_n$ multilayers [16], where AgSn indicates a Ag(6at.%Sn) alloy, gave behaviors that led to analysis by a 2CSR model, using the following equations.

### (B) 2CSR Model equations for simple $[F/N]_n$ multilayers.

The 2CSR model applied to an $[F/N]_n$ multilayer with superconducting leads predicts the following simple forms for $AR_{AP}$ and $A\Delta R$ [11,16]:

$$AR_{AP} = 2AR_{S/F} + n[\rho_N t_N + \rho^{*}_{F} t_F + 2 AR^{*}_{F/N}] \quad (1)$$

and

$$A\Delta R = n^2[\beta_F \rho^{*}_{F} t_F + 2\gamma_F AR^{*}_{F/N}]^2/AR_{AP} \quad (2)$$

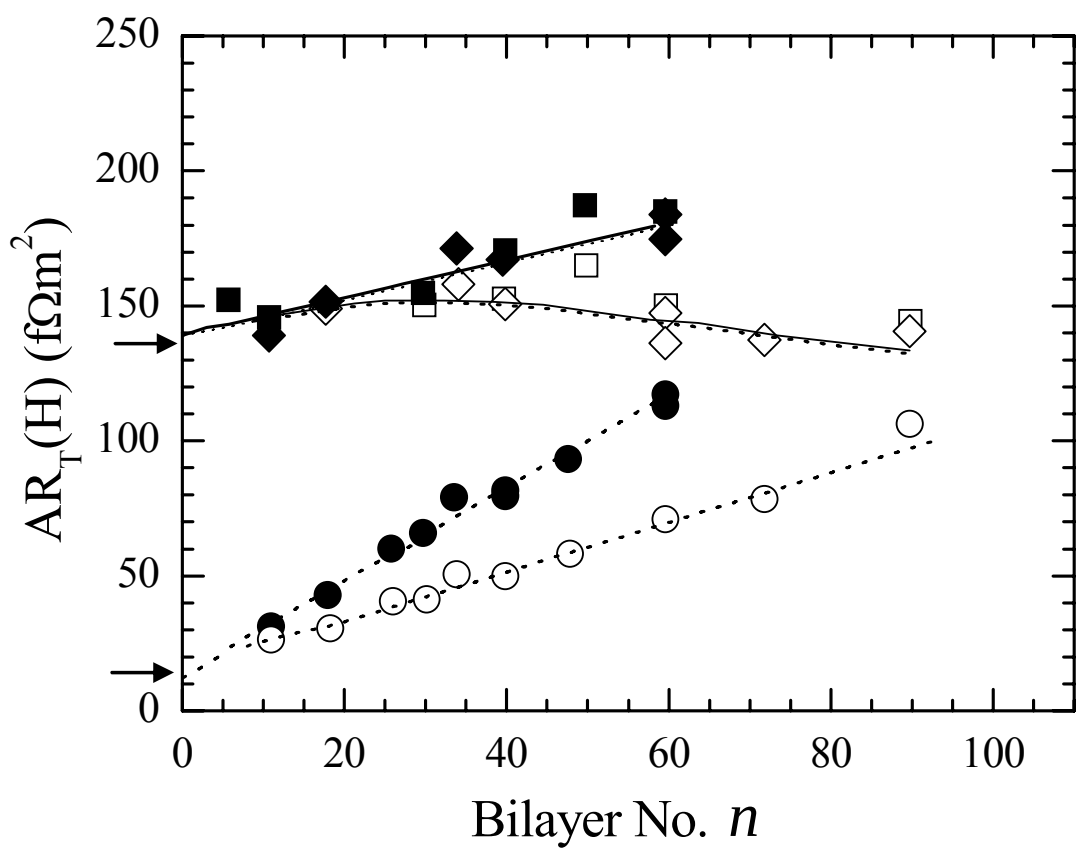

**Fig. 6.** Total AR ($AR_T$) vs bilayer number $n$ for samples of fixed total thickness 720 nm for sputtered $[Co(6)/Ag(t)]_n$ (circles) and $[Co(6)/AgSn(t)]_n$ (squares and diamonds indicate different sputtering runs) multilayers. Open symbols are for $AR_T(H_s) = AR_T$( minimum at the saturation field $H_s$) = $AR_P$ and filled symbols are for $AR_T(0) \approx AR_{AP}$. Some of the data have been corrected as explained in Pratt et al. [66]. The arrows indicate the independently predicted ordinate intercepts of ~ 13 $f\Omega m^2$ for Co/Ag and ~ 136 $f\Omega m^2$ for Co/AgSn. From: W.P. Pratt Jr. et al. [66].

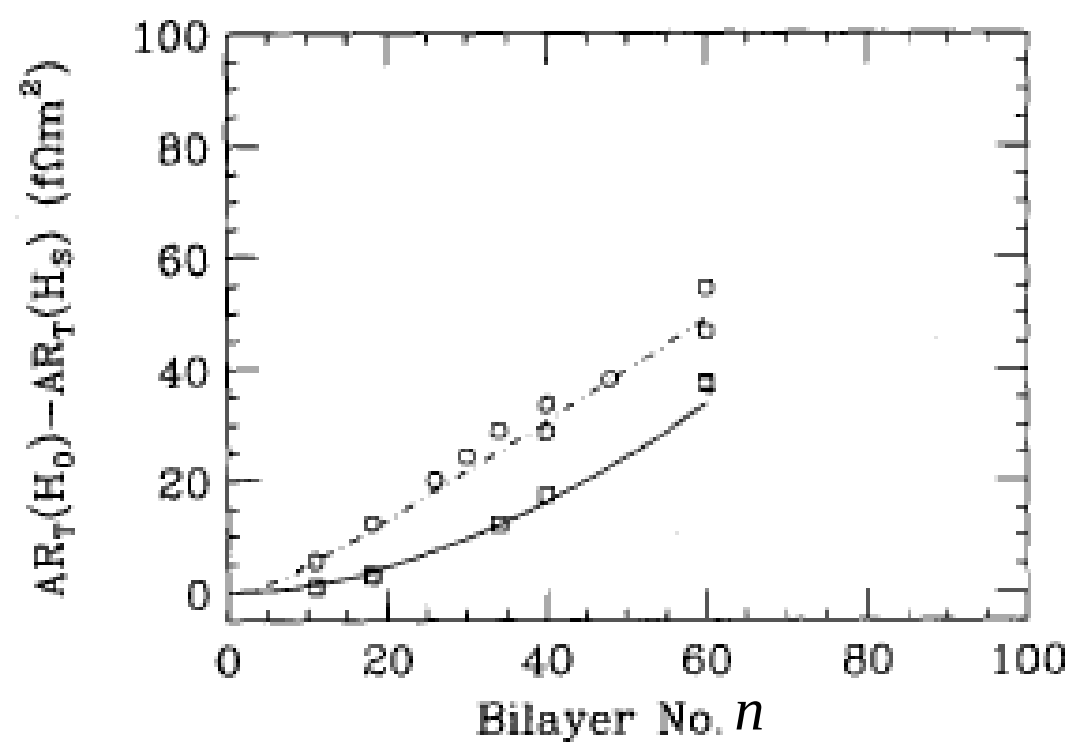

**Fig. 7**. $AR_T(0) - AR_T(H_s) \approx A\Delta R = AR_{AP} - AR_P$ vs. $n$ for some of the data in Fig. 6. Circles are for Co/Ag; squares are for Co/AgSn. From: S.F. Lee et al. [16]

Notice that the numerator of Eq. (2) depends only on properties of F and the F/N interfaces. Any constants in $AR_{AP}$, such as $\rho_N t_N$ or $2AR_{S/F}$, which are independent of F-moment orientations, do not contribute to the numerator of Eq. 2. For insight into the physics of Eq. (2), note that the product $\beta_F \rho_F^* = (\rho_F^{\downarrow} - \rho_F^{\uparrow})/4$ and the product $2\gamma_F AR_{F/N}^* = (AR_{F/N}^{\downarrow} - AR_{F/N}^{\uparrow})/2$.

Consider Eq. 1 for an $[F/N]_n$ multilayer with fixed $t_F$ and fixed total thickness $t_T = n(t_F + t_N)$. Eliminating the variable $t_N$, and neglecting the differences between $n$ and $n \pm 1$, gives [16]:

$$AR_{AP} = 2AR_{S/F} + \rho_N t_T + n[(\rho_F - \rho_N)t_F + 2\, AR_{F/N}^*]. \quad (1')$$

Eq. 1' then predicts that a plot of $AR_{AP}$ vs $n$ should give a straight line, with ordinate intercept $2AR_{S/F} + \rho_N t_T$ and fixed slope $[(\rho_F - \rho_N)t_F + 2\, AR_{F/N}^*]$. In contrast, for sufficiently small $n$, Eq. (2) predicts that $A\Delta R$ should first grow as $n^2$, and then transform to a linear variation as $n$ increases. The range of $n^2$ variation should increase as $\rho_N$ increases, thereby extending the range of $n$ over which the constant term in Eq. 1' remains dominant in the denominator of Eq. 2.

Lastly, multiplying both sides of Eq. 2 by $AR_{AP}$ and taking square roots gives [16,18]:

$$\sqrt{(AR_{AP})A\Delta R} = n[\beta_F \rho_F^* t_F + 2\gamma_F AR_{F/N}^*]. \quad (3)$$

Eq. (3) predicts that a plot of the square root on the left hand side vs $n$ should give a straight line passing through the origin, with a slope that is independent of $\rho_N$. That is, if the host metal N is alloyed with a small enough amount of a weakly spin-relaxing impurity, so that the spin-diffusion length, $l_{sf}^N$, remains long enough that the 2CSR model can still apply, the data for pure N and alloyed N should fall on exactly the same line, down to values of $n$ where the alloy layer thickness becomes comparable to the alloy $l_{sf}^N$.

To apply Eqs. 1-3 requires knowing $AR_{AP}$. For a simple $[Co/Ag]_n$ multilayer, Fig. 1 shows that the largest value of $AR_T$ (the total AR of the multilayer) occurs not at the coercive field, but rather in the initial, virgin state before any magnetic field is applied. We call this state $AR_T(0)$ and use it to approximate $AR_{AP}$, as justified in section IV. Using this state for each sample, and taking the smallest value of $AR_T$ (above the saturation field, $H_s$) as $AR_P$, allows the following tests of the 2CSR model.

**(C) Tests of the 2CSR model in $[Co/Ag]_n$ and $[Co/AgSn]_n$ multilayers.**

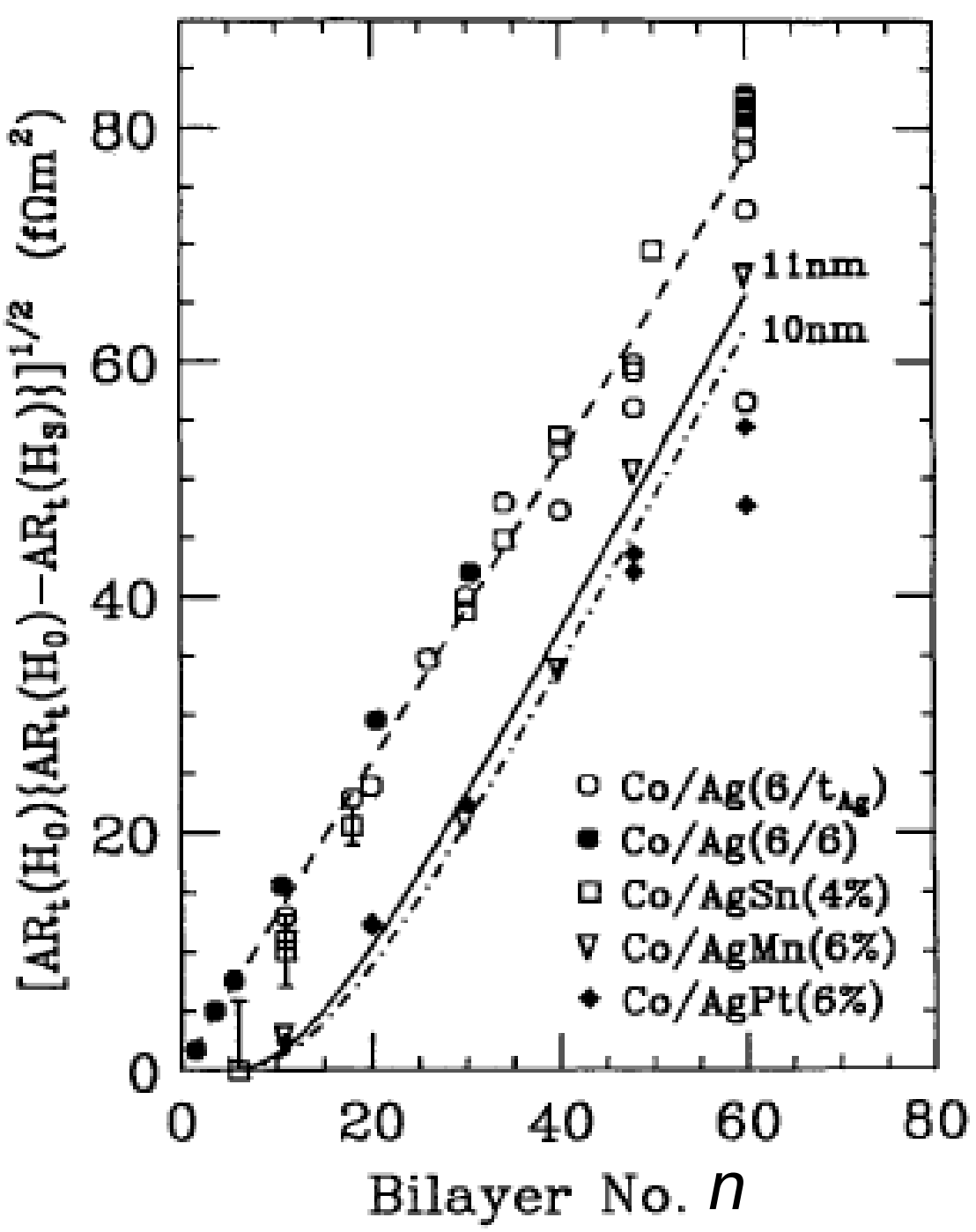

**Fig. 8**. Square root vs bilayer number *n* for a series of sputtered Co/Ag-based multilayers of fixed total thickness 720 nm, with 'pure' Ag and dilute Ag-based alloys.. Open circles are for $[Co(6)/Ag(t_{Ag})]_n$, with variable $t_{Ag}$ to keep the total thickness constant at 720 nm as *n* varies. Filled circles are for [Co(6)/Ag(6)]. As predicted by Eqn. (3), these two very different sets of data for $AR_T$ fall on the same straight line in the square root plot. Open squares are for $[Co(6)/Ag4\%Sn(t_{AgSn})]_n$ which, as predicted by Eq. (3), also falls on the same line (until the AgSn layer thickness approaches the AgSn spin-diffusion length). Open triangles are for $[Co(6)/Ag6\%Mn(t_{AgMn})]_n$ and filled diamonds are for $[Co(6)/Ag6\%Pt(t_{AgPt})]_n$, both of which have shorter spin-diffusion lengths than Ag or AgSn. The Dashed line is fit to the open circles. The solid curve is a VF fit to the Co/AgMn data with $l_{sf}^{AgMn} = 11$ nm. The broken curve is a VF fit to the Co/AgPt data with $l_{sf}^{AgPt} = 10$ nm. From J. Bass et al. [67].

Figs. 6 – 8 [16,66,67] show tests of these predictions for $[Co(6)/Ag(t_{Ag})]_n$ and $[Co(6)/Ag6\%Sn(t_{AgSn})]_n$ multilayers with fixed $t_T$ = 720 nm. Similar results were also obtained with $[Co/Cu]_n$ and $[Co/Cu4\%Ge]_n$ multilayers [67].

Fig. 6 (with the AgSn data slightly corrected as described in [66]) shows that the virgin state, total AR = $AR_T$ (0) assumed = $AR_{AP}$ is approximately linear in *n* for both cases, with very different ordinate intercepts due to the very different values of $\rho_{Ag}$ = 10 ± 1 nΩm and $\rho_{AgSn}$ = 185 ± 10 nΩm [16]. The arrows indicate the predicted ordinate intercepts assuming $2AR_{Nb/Co}$ = 6 ± 1 fΩm$^2$. The data are consistent with these predictions to within mutual uncertainties.

Fig. 7 [16] shows the $n^2$ variation of AΔR vs *n* for small *n*, with the range of $n^2$ variation being much larger for AgSn than for Ag.

Fig. 8 [67] shows that, despite the very different behaviors of $AR_{AP}$ in Fig. 6 and AΔR in Fig. 7, the square root data for Ag and AgSn obey the predictions of Eq. 3 of a single straight line passing through the origin, with the same slope for Ag and AgSn, down to small values of *n*, where the AgSn thickness becomes larger than its spin-diffusion length so that the 2CSR model no longer applies.

The data in Figs. 6-8 were taken as evidence in favor of the 2CSR model for Co/Ag and Co/AgSn. Similar behaviors of Co/Cu and Co/CuGe were taken as further evidence for the 2CSR model with F = Co [67]. Table 1 [18,19,20,67-71] and a more complete data collection in ref. [72], show that the values of $l_{sf}$ at 4.2K are long enough for the 2CSR model to apply to the Co, Ag, and AgSn layers in Figs. 6-8, at least down to small *n* for AgSn.

**(D) Test 2CSR and VF Parameters by predicting AΔR for Co/Cu/Py/Cu with no adjustment.**

As noted above, it was initially not obvious how closely $AR_o$ in Fig. 1 approximated $AR_{AP}$. This relationship was tested using $[Co(3)/Cu(20)/Py(8)/Cu(20)]_n$ hybrid spin-valves, chosen because the difference in coercive fields of Co(3) ($H_c \geq$ 100 Oe) and Py(8) ($H_c \leq$ 20 Oe) is large enough to give well-defined AP states. Fig. 5b [50] shows that this expectation is borne out. The first test involved predicting $AR_{AP}$ and $AR_P$ for these hybrid spin-valves using parameters for both $[Co/Cu]_n$ and $[Py/Co]_n$ multilayers derived from 2CSR model fits to the two sets of data, assuming $AR_o = AR_{AP}$. The predictions for $AR_{AP}$ and $AR_P$ were rather good [50], but the more challenging ones for AΔR were only fair (solid curves in Fig. 9 [50]). Soon afterwards, however, it was discovered that $l_{sf}^{Py}$ ~ 5.5 nm [73] was too short for the 2CSR model to be valid for $[Py/Cu]_n$ multilayers. The $[Py/Cu]_n$ data were refit with VF theory using $l_{sf}^{Py}$ ~ 5.5 nm. The dashed curves in Fig. 9 [74] show that the resulting no-free-parameter predictions for $[Co/Cu/Py/Cu]_n$ are improved. These agreements were taken to to jointly validate the VF model, the use of $AR_o = AR_{AP}$, and the separately derived value of $l_{sf}^{Py}$ ~ 5.5 nm.

**(E) Test of VF theory for Ag(X) & Cu(X) alloys with X = Pt, Mn, and Ni.**

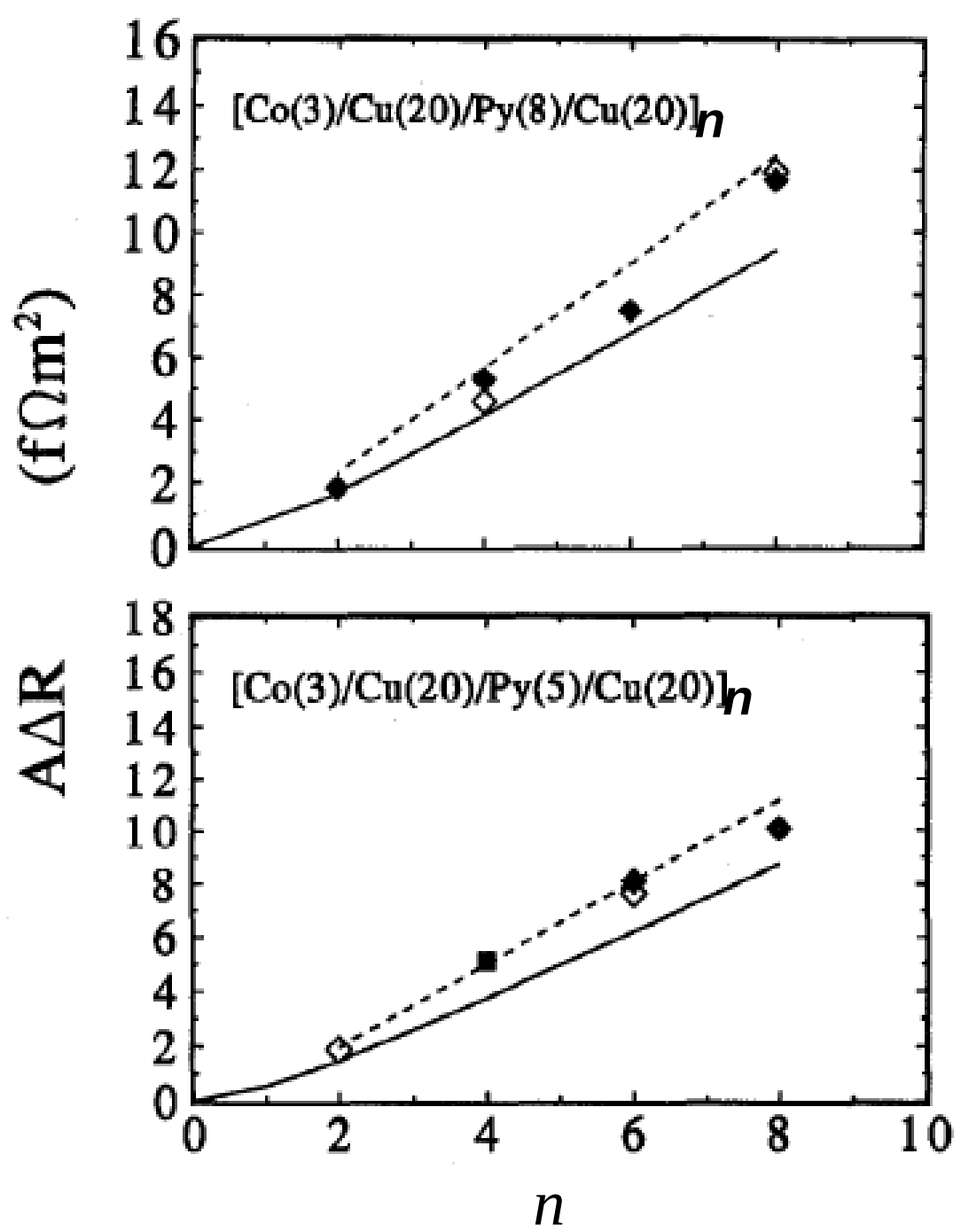

**Fig. 9**. AΔR vs bilayer number *n* for sputtered [Co(3)/Cu(20)/Py(5 or 8)/Cu(20)]$_n$ hybrid spin-valve multilayers. Solid curves are predictions from Py multilayer data analyzed using the 2CSR model (i.e., assuming $l_{sf}^{Py} = \infty$). Dashed curves are predictions from Py multilayer data analyzed using the VF model with $l_{sf}^{Py}$ = 5.5 nm. Filled and open symbols indicate different sputtering runs. From W.Pratt Jr. et al. [74].

The last early test involved applying VF theory beyond the 2CSR model to data for Ag- and Cu-based alloys with impurities that give stronger spin-relaxation, and thus shorter values of $l_{sf}^{N}$. The impurities were Pt and Mn in Ag as in Fig. 8 [67], and Pt, Mn, and Ni in Cu as in refs. [19,68,69,72]). VF theory gave values of $l_{sf}^{N}$ for these alloys, which were compared with independent predictions for Pt and Ni from conduction electron spin-resonance (CESR) measurements of spin-orbit cross-sections [71], and for Mn from calculations of spin-relaxation by spin-spin interactions [19]. The curves in Fig. 8 show the VF fits to the data for Pt and Mn in Ag. Table 1 shows the good agreement between the resulting values of $l_{sf}^{N}$ and the independent calculations. Further information about the CESR calculation is given in appendix I.

The positive results of the tests in sections VI.C, D, and E were taken as evidence of the validity of the VF and 2CSR models for analyzing CPP-MR data under appropriate conditions.

**(F) Problems with 2CSR model for 'separated' [Co/Ag]$_n$ and [Co/Cu]$_n$ multilayers.**

As explained in section IV, a reliable AP state can be obtained by making a hybrid spin-valve multilayer with two different F-layer thicknesses. If the 2CSR model is applicable to its constituents, then AΔR should be the same for two different forms of such a multilayer, e.g.: (a) 'interleaved' = [Co(1)/Cu(20)/Co(6)/Cu(20)]$_n$, and (b) 'separated' = [Co(1)/Cu(20)]$_8$[Co(6)/Cu(20)]$_8$, since AΔR from the 2CSR model is independent of the ordering of the single domain magnetizations of individual layers, so long as the overall magnetic order is P at large field and AP at an appropriate intermediate field. The solid symbols in Fig. 5c show AR(H) for such Co-based multilayers. Their AΔRs are very different. The 2CSR model works well for the interleaved sample, but doesn't work for the separated sample. The best available spin-diffusion length in Co ($l_{sf}^{Co}$ ~ 60 nm [23,75]) is too long to explain the observed difference [76,77]. The difference has been attributed to mean-free-path effects—including ballistic transport [76,78,79], or to spin-relaxation at the Co/Cu interfaces (i.e., to $\delta_{Co/Cu}$) [21,77]. For details see Appendix C of [72]. While we favor interfacial spin-relaxation, the references and discussion in Appendix C should let the reader form his/her own opinion.

## VII. Bulk CPP-MR Parameters, $\beta_F$, $l_{sf}^{F}$, and $l_{sf}^{N}$.

The first sets of bulk and interface parameters were derived together for simple [Co/Ag]$_n$ and [Co/Cu]$_n$ multilayers assuming applicability of a 2CSR model—i.e., no spin-relaxation. Subsequently, values of $\beta_F$ and $l_{sf}^{F}$ for F-based alloys have also been estimated together using hybrid spin-valves or EBSVs. Values of $l_{sf}^{N}$ have been derived by inserting layers of N in the middle of Py-based EBSVs. In this section we describe these techniques and present what we believe to be the most reliable values of bulk parameters. We assume that values of the other two bulk parameters, $\rho_F$ and $\rho_N$, are measured separately on samples deposited in the same way as the multilayers, as discussed in section II.

### A) Bulk anisotropy parameter, $\beta_F$.

**Table 1. Selected alloy values of spin-diffusion lengths, $l_{sf}^{N}$ (nm), from CPP-MR and Conduction Electron Spin Resonance (CESR).** Also listed are the sample type (multilayer (ML) or spin-valve (SV)), and alloy residual resistivity, $\rho_o$.

| Alloy | Tech. | $l_{sf}^{N}$ (CPP-MR) | $l_{sf}^{N}$ (CESR) [19,71] | $\rho_o$(nΩm) | Ref. |
|---|---|---|---|---|---|
| Ag(4%Sn) | ML | ≈ 39 | | 200±20 | [18,68] |
| Ag(6%Pt) | ML | ≈ 10 | ≈ 7 | 110±20 | [18] |
| Ag(6%Mn) | ML | ≈ 11 | ≈ 12* | 110±25 | [18] |
| Cu(4%Ge) | ML | ≥ 50 | ≈ 50 | 182±20 | [67,68]; |
| Cu(6%Pt) | ML | ≈ 8 | ≈ 7 | 130±10 | [18] |
| Cu(6%Pt) | SV | 11±3 | ≈ 7 | 160±30 | [20] |
| Cu(7%Mn) | ML | ≈ 2.8 | 3±1.5* | 270±30 | [18] |
| Cu(22.7%Ni) | ML | 7.5 | 6.9 | 355 | [69] |
| Cu(22.7%Ni) | SV | 8.2 ± 0.6 | 7.4 | 310±20 | [70] |

**Table 2. $\beta_F$ at 4.2Kfor dilute F-based alloys from DMR or CPP-MR with effects of finite $l_{sf}^{F}$.**

| F-host | Impurity | $\beta_F$ (DMR)[8] | $\beta_F$(CPP-MR) | $\beta_F$(Calc.)[84] |
|---|---|---|---|---|
| Ni | Fe | +0.88 ± 0.1 | +0.73 ± 0.1 [80] | +0.68 |
| Ni | Fe | | +0.76 ± 0.07 [81] | +0.68 |
| Ni | Cr | -0.54 ± 0.1 | -0.35 ± 0.1 [82] | |
| Co | Fe | +0.85 ± 0.1 | +0.65 ± 0.05 [83] | |

As explained in section II, values of $\beta_F$ are unique only for dilute alloys containing a known concentration of a known impurity. We therefore start with values of $\beta_F$ for such alloys, and then turn to values for nominally 'pure' metals.

**(A1) $\beta_F$ for F-alloys.**

Table 2 [80-84] compares values of $\beta_F$ for some F-based alloys derived from CPP-MR measurements using the full VF theory, including finite values of $l_{sf}$, with values derived from studies of Deviations from Matthiessen's Rule [8], and with a recent no-free-parameter calculation for Py = $Ni_{80}Fe_{20}$[84]. The agreements of the CPP-MR results with both Deviations from Matthiessen's Rule and the calculation are generally satisfactory. For more complete tables, including less reliable values derived from a 2CSR model neglecting finite $l_{sf}^{F}$, see refs. [7,85,86].

As an example of determining the values of $\beta_F$ and $l_{sf}^{F}$ in Table 2 for a specific alloy, we use a dilute $Ni_{97}Cr_3$ alloy (hereafter just NiCr ) that has a negative $\beta_{NiCr}$ but a positive $\gamma_{Ni/Cu}$. Combining NiCr in an N = Cu hybrid spin-valve with Py, for which both $\beta_{Py}$ and $\gamma_{Py/Cu}$ are positive, causes the CPP-MR to change sign as $t_{NiCr}$ is increased, thereby changing the dominant scattering associated with NiCr from positive when scattering from the NiCr/Cu interface dominates, to negative when scattering from the NiCr bulk dominates. 3 parameters: $\beta_{NiCr}$, $l_{sf}^{NiCr}$, and the product $\gamma_{NiCr/Cu} AR^*_{Ni/Cr/Cu}$, were fit with VF theory [82] to two independent sets of samples: (a) hybrid spin-valves of the form $[Py(6)/Cu(20)/NiCr(t_{NiCr})/Cu(20)]_{10}$, and (b) EBSVs of the form $[FeMn(8)/NiCr(t_{NiCr})/Cu(20)/Py(6)]$. All of the other parameters were fixed at previously measured values. The dashed and solid curves in Fig. 10 [82] are alternative fits to the hybrid SV data alone with the parameters listed. Combining the hybrid and EBSV results gave 'best values' of $\beta_{NiCr}$ = - 0.35 ± 0.1, $l_{sf}^{NiCr}$ = 3 ± 1 nm, and $\gamma_{NiCr/Cu} AR^*_{Ni/Cr/Cu}$ = 0.16 ± 0.07 fΩm$^2$. Table 2 shows that this $\beta_{NiCr}$ is comparable to the value $\beta_{NiCr} = -0.54^{+0.1}_{-0.15}$ ) found from DMR studies [8]. In contrast, the $\beta_{NiCr}$ = - 0.13± 0.01 [85] inferred from a fit with only a simple 2CSR model (i.e., assuming $l_{sf}^{NiCr} = \infty$) is too small.

**(A2) $\beta_F$ for 'pure' F-metals.**

Table 3 [4,47,74,81,87-91] compares values of $\beta_{Co}$, $\beta_{Py}$, and various other parameters for Co/Cu and Py/Cu interfaces derived in different laboratories using different techniques. Note that the Co layers have quite different

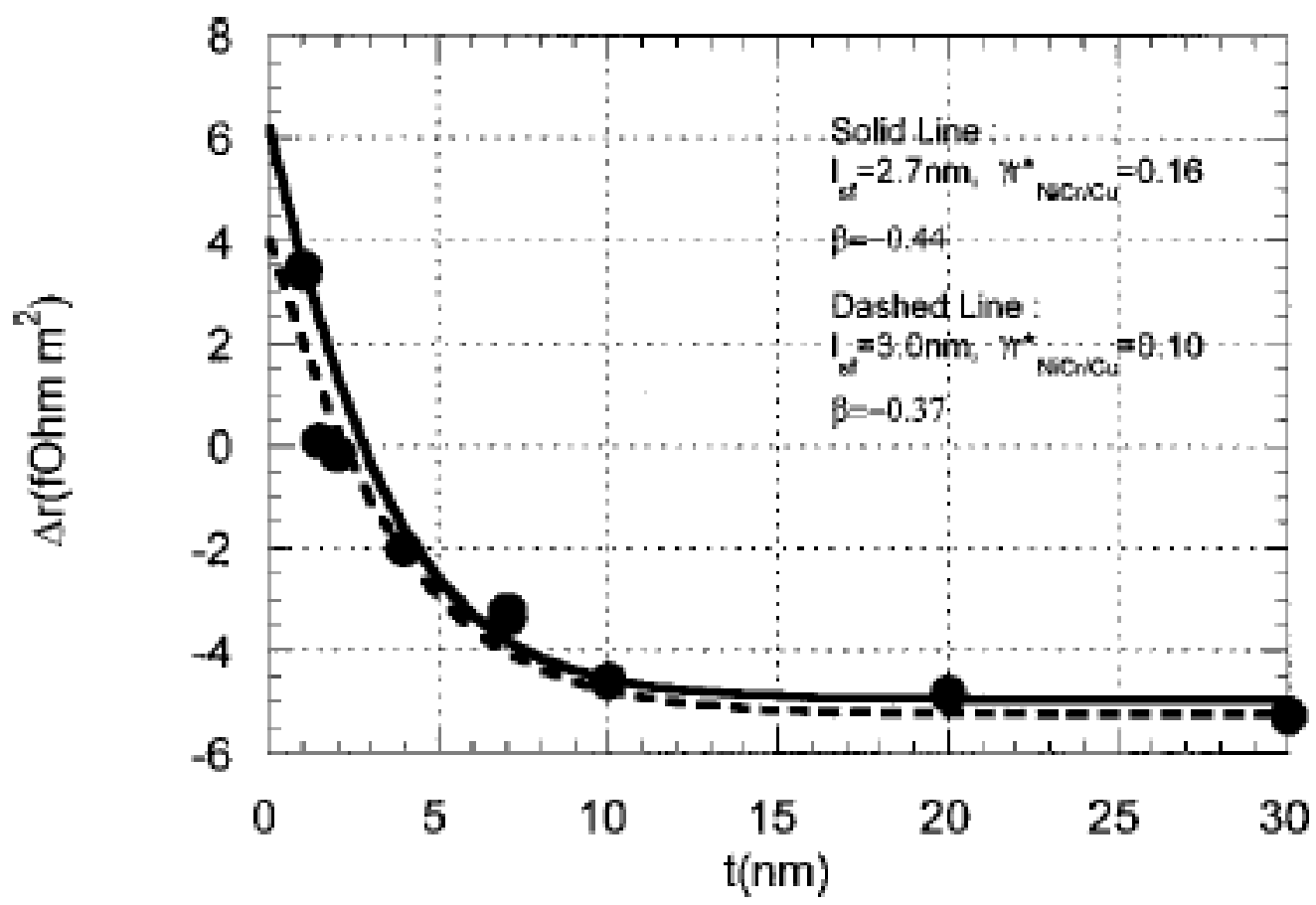

Fig. 10. AΔR vs NiCr thickness t for sputtered hybrid spin-valves of the form Py/Cu/NiCr(t)/Cu. The solid and dashed curves are fits with different assumed parameters showing the range of uncertainty. From W. Park et al., [82]

residual resistivities (i.e., different nominal purities). The values of $\beta_{Co}$ are more similar than might have been expected from the cautionary remarks above. Values of $\beta_F$ for Fe [92] and for Ni [93] have been derived by too few groups to check for consistency.

The parameters for Co/Cu given in Table 3 were derived with both superconducting leads and nanowires by applying a simple 2CSR model to sets of measurements of $AR_{AP}$, $AR_P$, AΔR, and CPP-MR on multilayers with different thicknesses and number of Co and Ag or Co layers. For examples of the procedures used see the references listed in Table 3. The studies with superconducting leads used Eqs. 1-3 and equivalent forms for $[Co/N]_n$ multilayers with different combinations of fixed and varied thicknesses of Co and N = Cu and different values of *n*. The studies with nanowires used equivalent equations for the CPP-MR, such as

$$[A\Delta R/AR_{AP}]^{-1/2} = (\rho^*_F t_F + 2AR^*_{F/N})/(\beta_F \rho^*_F t_F + \gamma_{F/N} 2AR^*_{F/N}) + \rho_N t_N/((\beta_F \rho^*_F t_F + \gamma_{F/N} 2AR^*_{F/N}), \quad (4)$$

with

$$AR_{AP} = n[\rho^*_F t_F + 2AR^*_{F/N} + \rho_N t_N]. \quad (5)$$

**Table 3. Comparing Parameters for Co/Cu and Py/Cu from different groups.**
MSU and Leeds values found with $R_o$. Louvain-Orsay (LO) and Lausanne (Laus.) were found with $R_{Pk}$. Eindhoven (Eind.) were extrapolated from grooved samples.

| Parameter | MSU Sup.Lead,4.2 K [4,74,81,87] | Leeds Sup.Leads 4.2K [88] | LO Nanowires 77K [89] | Laus. Nanowires, 20K [90,91] | Eind. Grooved, 4.2K [47] | Calc. [84]. |
|---|---|---|---|---|---|---|
| $\rho_{Cu}$(nΩm) | 6 ± 1 | 13 ± 3 | 31 | 13-33 | 3.6 | |
| $\rho^*_{Co}$ (nΩm) | 75 ± 5 | 30 ±6 | 180 ± 20 | 510-570 | 57 | |
| $\beta_{Co}$ | 0.46 ± 0.05 | 0.48 ± 0.04 | 0.36 ± 0.02 | 0.46 ± 0.05 | 0.27 | |
| $\gamma_{Co/Cu}$ | 0.77 ± 0.04 | 0.71 ± 0.02 | 0.85 ± 0.15 | 0.55 ± 0.7 | 0.52 | |
| $2AR^*_{Co/Cu}$ (fΩm$^2$) | 1.02 ± 0.04 | 0.86 ± 0.08 | | 0.6-2.2 | 0.4 | |
| $l^{Co}_{sf}$ (nm) | ≥ 40 | | 59 ± 18 | | | |
| | | | | | | |
| $\rho^*_{Py}$ | 291 ± 90 | | 263 | | | |
| $\beta_{Py}$ | 0.76 ± 0.07 | | 0.8 ± 0.1 | | | 0.68 |
| $\gamma_{Py/Cu}$ | 0.7 ± 0.1 | | 0.8 ± 0.1 | | | |
| $2AR^*_{Py/Cu}$ | 1.00 ± 0.08 | | | | | |
| $l^{Py}_{sf}$ (nm) | 5.5 ± 1 | | 4.3 ±1 | | | 5.5 |

Eq. 4 had to be used instead of Eq. (3) because the effective area A of the parallel collection of an unknown number of nanowires was unknown. Eq. (5) is included to make clear that the absence of superconducting leads, plus the several micron long nanowire with a large number *n* of repeats, together should allow neglect of the lead resistance.

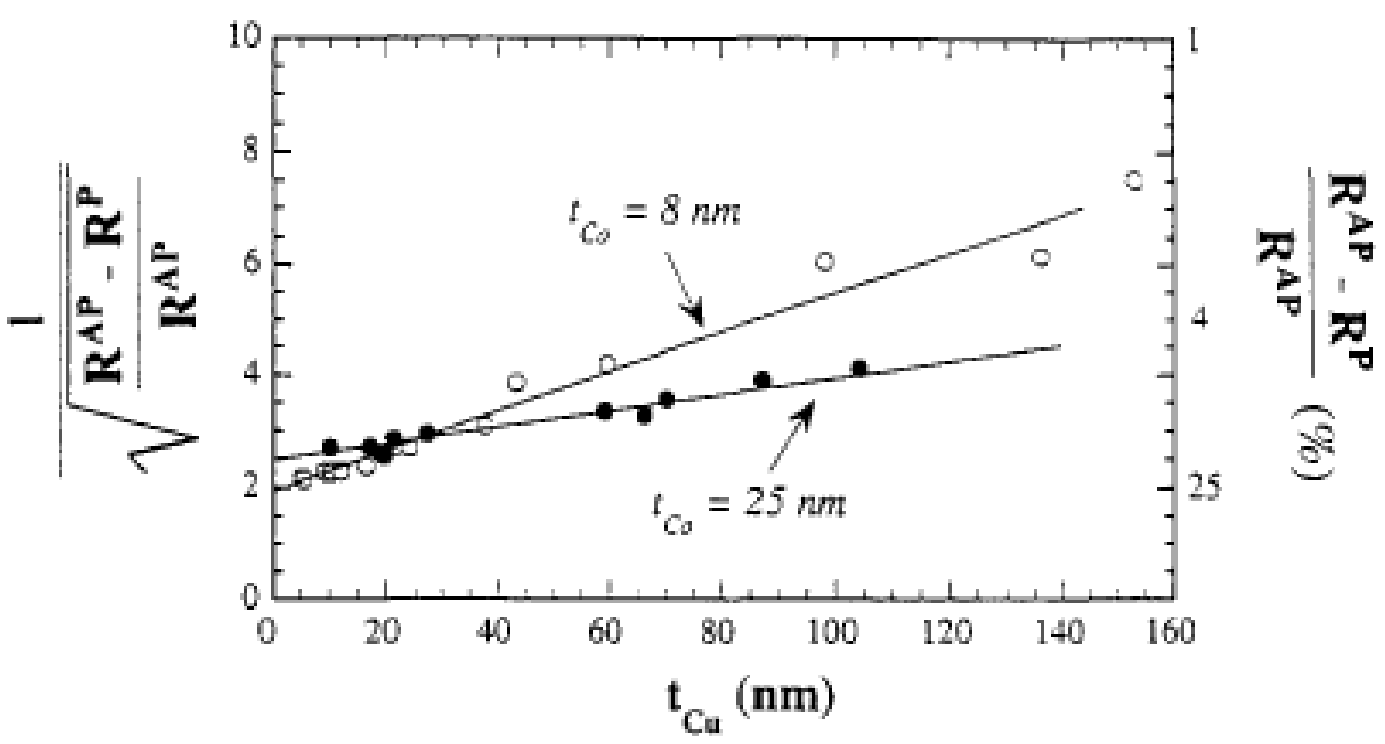

**Fig. 11**. 1/(square root) vs $t_{Cu}$ for nanowire Co/Cu multilayers with two fixed thicknesses of Co. From Eq. 4, the ordinate value at the crossing point of the two lines should equal $1/\beta_{Co}$. From: L. Pireaux et al. [91]

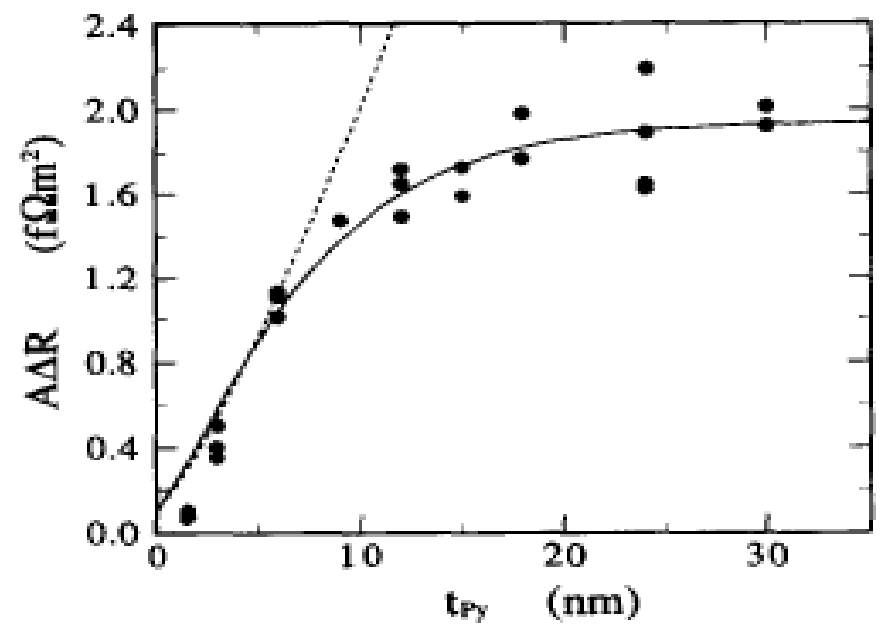

**Fig. 12**. AΔR vs $t_{Py}$ for sputtered symmetric Py-based EBSVs of the form FeMn(8)/Py($t_{Py}$)/Cu(20)/Py($t_{Py}$). The solid curve is a fit with $l_{sf}^{Py}$ = 5.5 nm. The dotted

From Eq. (4), a plot of $[A\Delta R/AR_{AP}]^{-1/2}$ vs $t_N$ for fixed $t_F$ should give a straight line, and the lines for different values of $t_F$ should all cross at a point with vertical coordinate equal to $(1/\beta_F)$. Fig. 11 [91] shows an example of such behavior at 77K with $t_F < l_{sf}^F$ and $t_N < l_{sf}^N$.

**(B) Spin-Diffusion lengths**.

Values of $l_{sf}^N$ and $l_{sf}^F$ found by a variety of techniques, including CPP-MR, are collected in [72]. We describe here how CPP-MR measurements give $l_{sf}^N$ and $l_{sf}^F$. As with $\beta_F$ above, $l_{sf}$ only has a unique value in an alloy with a known concentration of a single dominant scatterer.

For convenience in using equations, we begin with $l_{sf}^F$.

**(B.1)** $l_{sf}^F$.

The first CPP-MR measurement of a short $l_{sf}^F$ was made for F = Py. The analysis involved applying VF theory to the data in Fig. 12 [73] of AΔR vs Py-thickness, $t_{Py}$, for Py-based symmetric EBSVs (equal thicknesses of Py) of the form Nb/Cu/FeMn(8)/Py($t_{Py}$)/Cu(20)/Py($t_{Py}$)/Cu/Nb. AΔR first increases approximately linearly with increasing $t_{Py}$, and then bends over and saturates at a constant value for $t_{Py} >> l_{sf}^{Py}$. The solid curve is a numerical VF fit with $l_{sf}^{Py}$ = 5.5 nm and the complete set of other parameters given in [74]. For comparison, the dashed curve is the VF prediction for the same parameters, except with $l_{sf}^{Py} = \infty$. Note that the solid curve first rises slightly above the dashed curve, and then bends over and eventually becomes constant (saturates). These behaviors can be understood by considering the VF equation for AΔR in the limit $t_{Py} >> l_{sf}^{Py}$ = 5.5 nm:

$$A\Delta R = 4[\beta_{Py}\,\rho^*_{Py}\,l_{sf}^{Py} + 2\gamma_{Py}\,AR^*_{Py/Cu}]^2/(2\,\rho^*_{Py}\,l_{sf}^{Py} + 2\,AR^*_{Py/Cu} + \rho_{Cu}t_{Cu}). \quad (6)$$

Comparing Eq. 6 with Eq. 2, we see that the numerators are almost the same, but the denominators are very different. The only change in the numerator is that $t_{py}$ in Eq. 2 is replaced by $l_{sf}^{Py}$ in Eq. 6. Once $t_{Py}$ becomes significantly larger than $l_{sf}^{Py}$, the numerator no longer increases with increasing $t_{Py}$. In contrast, for the full EBSV, the denominator in Eq. 2 reduces in Eq. 6 to only the ‘active part’ of the EBSV—i.e., the central part of the EBSV bounded by the distances $l_{sf}^{Py}$ outside of each of the Py/Cu interfaces. The contributions to the denominator from the FeMn layer and the S/Py boundaries that appear in Eq. 2 have disappeared from Eq. 6. It is the elimination of these ‘outer resistances’ that leads to the increase of the solid curve over the dashed curve in the region $t_{Py} \leq l_{sf}^{Py}$ = 5.5 nm. This elimination both simplifies the calculation of the constant AΔR in the long $t_{Py}$ limit, and also allows AΔR to grow larger than it would have if the denominator had remained the full $AR_{AP}$.

The short value of $l_{sf}^{Py}$ derived in [73] using crossed Nb strips, was confirmed by measurements on nanowires in [94], and by a no-free-parameter calculation [84].

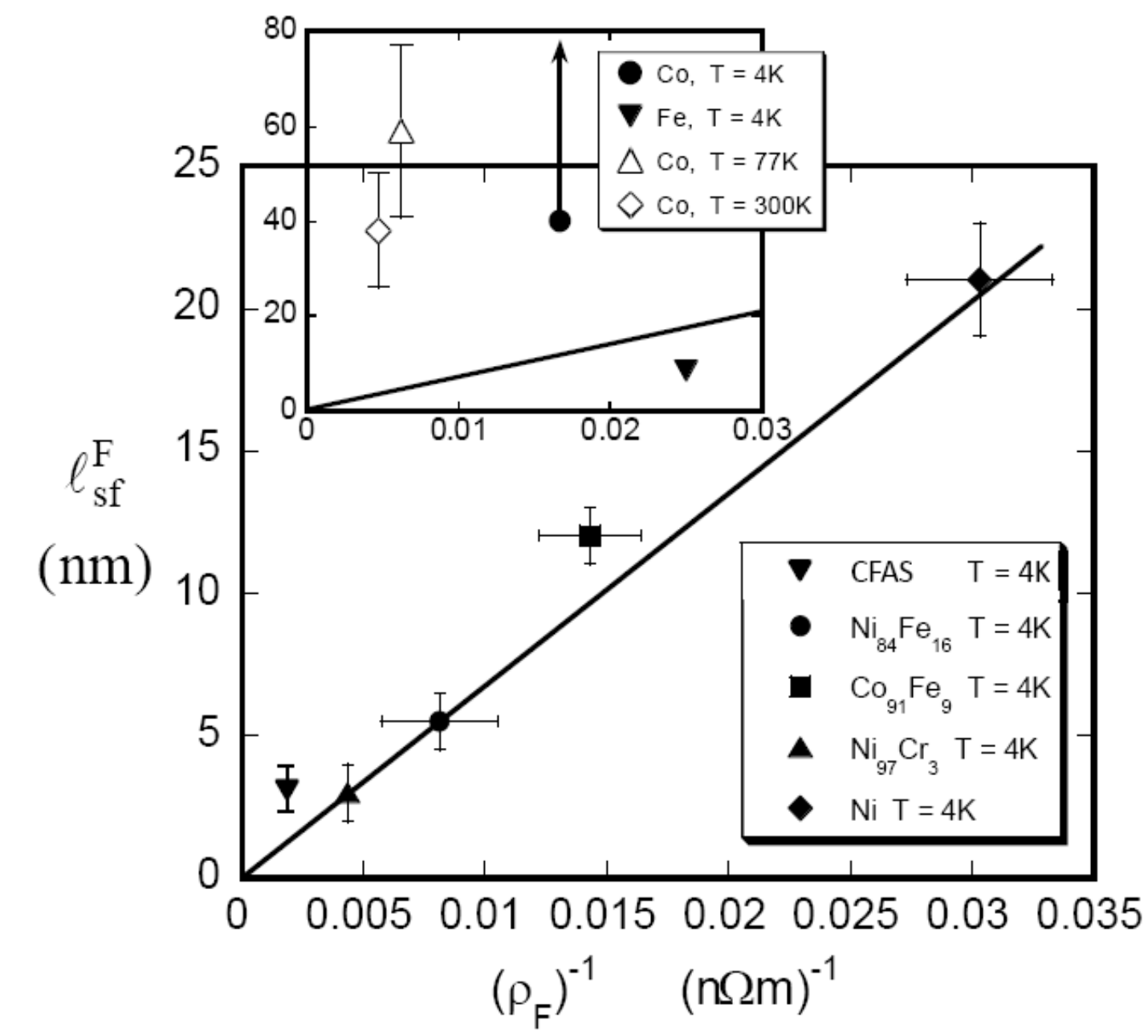

**Fig. 13. Spin-diffusion length,** , vs inverse residual resistivity, $1/\rho_F$ , for a series of sputtered F-metals and F-based alloys. The symbols are explained on the figure. CFAS = $Co_2FeSi_{0.5}Al_{0.5}$. The line is a best fit, going through 0,0, to the data points in the main graph, but not including the one for CFAS. The references are. In inset: △ [23]; [23]; ● [75]; ▼ [92]; In main figure: ◆ [93]; ▲ [82]; ■ [83]; ● [73]; ▼ [95]. After J. Bass and W.P. Pratt Jr. [72].

For a given dilute alloy, both λ and $l_{sf}$ should be inversely proportional to the impurity concentration. In a free electron model, λ at 4.2K is also inversely proportional to the residual resistivity, $\rho_o$. Thus, one should be able to at least roughly compare the magnitudes of spin-relaxation in dilute alloys and nominally pure F-metals by plotting $l_{sf}$ vs $(1/\rho_o)$. Fig. 13 [72] shows such a plot for several F-metals and alloys [23,73,75,82,83,92-95]. Most of the data fall close to a single straight line, with Co the most significant outlier. The first evidence that the spin-diffusion length in nominally pure Co, $l_{sf}^{Co}$ ~ 60 nm at 77K and ~ 40 nm at 300K, might be unusually long was found using nanowires in [23,91], and later confirmed with crossed Nb strips [75]. Fig. 14 [83] compares the behaviors of AΔR vs $t_F$ for F = $Co_{91}Fe_9$ and Co at 4.2K. In the growth of AΔR, the effect of longer $l_{sf}^{F}$ ($l_{sf}^{Co}$ ~ 60 nm [23,91] > $l_{sf}^{CoFe}$ ~ 12 nm [83]), is outweighed by the effect of larger $\beta_F$ ($\beta_{CoFe}$ ~ 0.66 [83] > $\beta_{Co}$ ~ 0.46 [4]).

**(B.2a) $l_{sf}^{N}$ for Alloys.**

Two techniques have been used to find $l_{sf}^{N}$ for alloys from CPP-MR measurements, both using crossed Nb strips. The first used simple $[Co/AgX]_n$ and $[Cu/CuX]_n$ multilayers, where X indicates a dilute impurity, and applied VF theory to deviations from Eq. 3 when the N = AgX or CuX layer thickness becomes longer than $l_{sf}^{N}$. The results are collected in [72], and examples are shown in Fig. 8. The second is described in section B.2b. Table 1 shows that the derived values of $l_{sf}^{N}$ from the two techniques agree to within mutual uncertainties, and also agree well with either values calculated from conduction electron spin-resonance (CESR) measurements of spin-flipping cross-sections, or calculated effects of spin-spin scattering). Supplementary note #2 explains how different lengths are obtained and how they are to be compared.

**(B.2b) $l_{sf}^{N}$ for Nominally Pure Metals.**

Along with a variety of other methods [72], the second technique has been used to also find values of $l_{sf}^{N}$ for nominally pure N-metals. It involves sandwiching the N-metal of interest in the middle of a Py-based EBSV of the form Nb(250)/Cu(10)/FeMn(8)/Py(24)/Cu(10)/N($t_N$)/Cu(10)/Py(24)/Cu(10)/Nb(250), measuring AΔR as a function of the N-layer thickness, $t_N$, and analyzing the data using the theory of VF. The published values of $l_{sf}^{N}$ are collected in [72]. Fig. 15 [20] shows examples of such data as plots of log (AΔR) vs $t_N$ , with the resulting values of $l_{sf}^{N}$ given in the caption. To explain these data, Eq. (4) must be generalized to include the effects of spin-relaxation associated with inserting the N-layer and its two N/Cu interfaces. The data for CuPt = Cu(6%Pt) were taken to compare with the values found for CuPt by the first technique described just above. Overlap of the new value, $l_{sf}^{CuPt}$ = 11 ± 3 nm, and the older value of $l_{sf}^{CuPt}$ ≈ 8 nm from ref. [18], tends to validate both techniques.

Even if there is no spin-relaxation within N, inserting a thickness $t_N$ of N into the middle of the central Cu layer of a Py-based EBSV adds to the denominator of Eq. 6 two terms, one from the bulk of N, $\rho_N t_N$, and one from the two N/Cu interfaces. Sputtered samples typically have interfaces that intermix over 3-4 monolayers (ML)— equivalent to 0.6-0.9 nm [24]. The initial rapid decreases of AΔR in Fig. 15 with increasing $t_N$ are probably due primarily to formation of the interfaces, and to spin-relaxation at them. Support for this argument comes from: (a) the absence of any such decrease for CuPt, where a significant 'interface' is not expected between Cu and Cu(6%Pt); (b) the near absence of any decrease for Ag, since the Cu/Ag interface resistance is very small; and (c) the largest decrease being for W, where both $AR_{Cu/W}$ and any spin-relaxation at the Cu/W interface (see section VIII.D below) are largest. The values of $AR_{Cu/N}$ are given in the caption to Fig. 15.

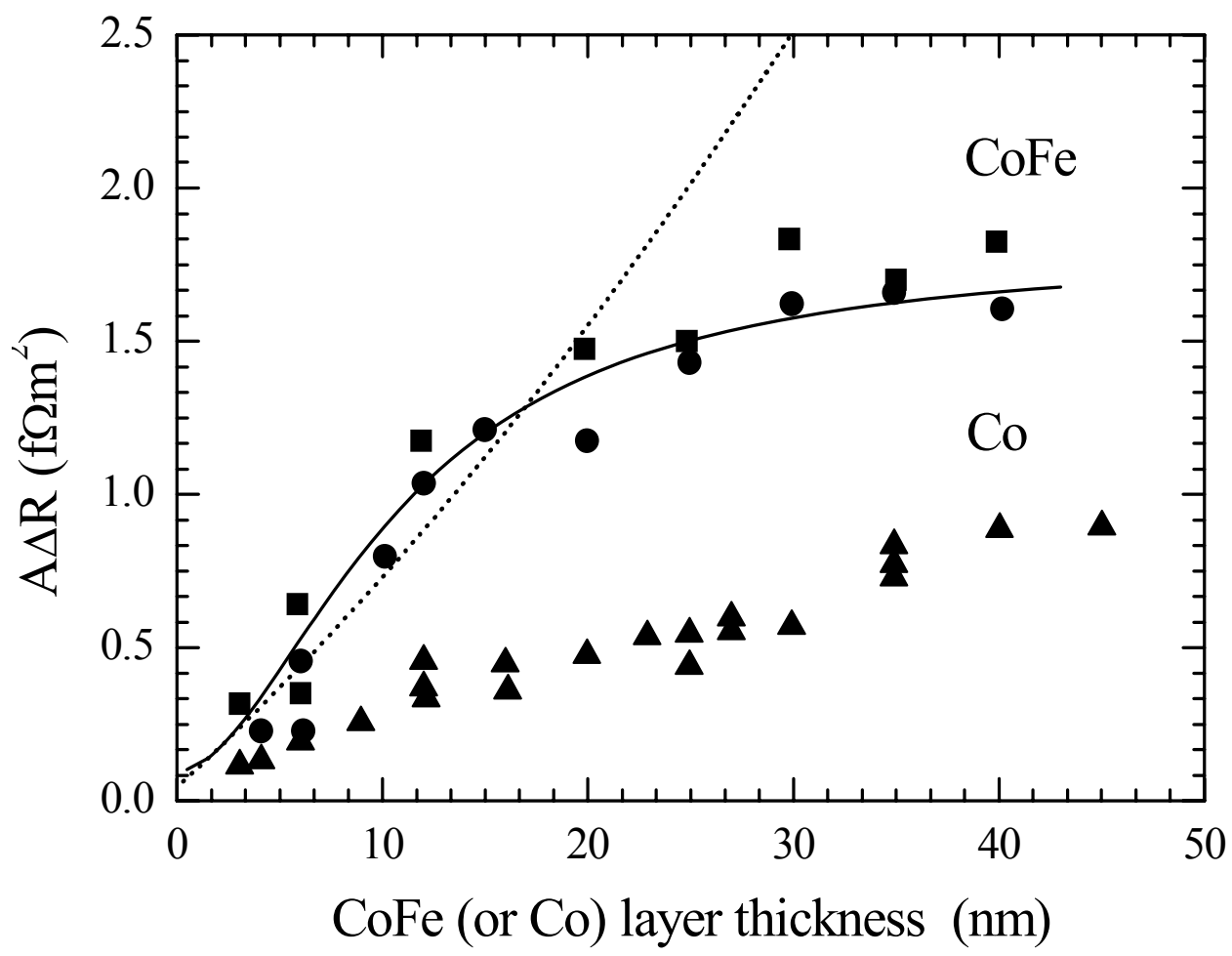

**Fig. 14.** AΔR vs thickness $t_F$ for F = Co (filled triangles for several sputtering runs) or F = $Co_{91}Fe_9$ = CoFe (filled circles or squares are for two different sputtering runs)**.** The solid curve through the CoFe data is a fit with $\beta_F = 0.66$ and $l_{sf}^{CoFe}$ = 12 nm. The dotted curve is VF theory with $\beta_F = 0.66$ and $l_{sf}^{CoFe} = \infty$. The Co data are consistent with $\beta_F = 0.46$ and $l_{sf}^{Co} \geq 20$ nm. The larger AΔR for CoFe is due mainly to the larger $\beta_{CoFe}$. From A. Reilly et al. [83].

Once the two interfaces are fully formed, and the denominator of Eq. 4 has been increased by $\rho_N t_N + 2AR_{Cu/N}$, the denominator stays constant, and the additional logarithmic decrease in AΔR with increasing $t_N$ is attributed to spin-relaxation within N—i.e., to finite $l_{sf}^{N}$, as explained in [20]. The dotted and dashed curves in Fig. 15 are VF theory fits to the data with the values of $l_{sf}^{N}$ given in the caption.

The decrease in AΔR with increasing layer thickness of the disordered sputtered antiferromagnet (AF) FeMn is so fast that it was attributed to strong spin-relaxation at the FeMn/Cu interface [20]. Similar behavior was recently reported for the disordered sputtered AF IrMn [96], along with evidence that spin-relaxation in the bulk of these AFs is probably also strong.

**VIII. Interface Parameters: $\gamma_{F/N}$, $2AR^*_{F/N}$, $2AR_{N1/N2}$, $2AR_{S/F}$, and δ.**

We first discuss $\gamma_{F/N}$ and $2AR^*_{F/N}$, then $2AR_{N1/N2}$, and finally $\delta_{N1/N2}$; $\delta_{F/N}$; and $\delta_{F1/F2}$. An extensive table of values of $\gamma_{F/N}$ is given in ref. [7]. Tables of $2AR^*_{F/N}$ and $2AR_{N1/N2}$ are given in refs. [7] and [97].

**Table 4. Selected examples of the product $\gamma_{F/N}2AR^*_{F/N}$, and its constituents $\gamma_{F/N}$ and $2AR^*_{F/N}$.** The values of $\gamma_{F/N}2AR^*_{F/N}$ are rounded to 1 significant figure. All derivations neglect interface spin-flips.

| **Metal Pair** | $\gamma_{F/N}$, | $2AR^*_{F/N}$(fΩm²) | $\gamma_{F/N}2AR^*_{F/N}$(fΩm²) |
|---|---|---|---|
| Co/Cu | 0.87 | 1.0 | 0.9 [4] |
| Co/Ag | 0.85 | 1.1 | 0.9 [26] |
| Fe/Cr | -0.7; -0.59 | 1.6 | 1 [98,99] |
| Fe/Cu | 0.55 | 1.5 | 1 [92] |
| Py/Cu | 0.7 | 1.0 | 0.7 [74,81] |
| Ni/Cu | 0.3 | 0.36 | 0.1 [93] |
| Py/Al | 0.025 | 8.5 | 0.2 [100] |
| $Co_{90}Fe_{10}$/Al | 0.1 | 10.6 | 1 [100] |
| Fe/Al | 0.05 | 8.4 | 0.4 [100] |

**(A) Interface Anisotropy Parameter, $\gamma_{F/N}$, and Enhanced Specific Resistance, $2AR^*_{F/N}$.**

The parameters $2AR^*_{F/N}$ and $\gamma_{F/N}$ are usually determined together by fits to multilayer or EBSV data. Eqs. (2) and (4) indicate that the interfacial quantity that best determines AΔR is their product $\gamma_{F/N}2AR^*_{F/N}$. In Table 4 [4,26,74,81,92,93,98-100] we list values of $\gamma_{F/N}$, $2AR^*_{F/N}$, and $\gamma_{F/N}2AR^*_{F/N}$ for a selection of F/N pairs to show how the three quantities vary. A more complete list of values of $\gamma_{F/N}$ is given in ref. [7]. To focus on the differences in magnitude of $\gamma_{F/N}\, 2AR^*_{F/N}$, we round its values in Table 4 to one significant figure. The largest values are all of order unity. For a dilute F-alloy we expect similar values to those for the host F-metal, since the interface consists mainly of the F- and N-atoms. The only direct test so far gave close agreement for Co and $Co_{90}Fe_{10}$ [101]. Comparing values of $\gamma_{F/N}$ and $2AR^*_{F/N}$ for Co/Cu and Py/Cu in Table 3 from different laboratories shows reasonable agreement for $\gamma_{Co/Cu}$ with superconducting leads and one nanowire result, but not so good for the other nanowire result or with grooved substrates. Table 5 [4,24,60-64,98,102-104] shows that the values of $2AR^*_{F/N}$ for the lattice matched F/N pairs Co/Cu and Fe/Cr, and the F1/F2 pair Co/Ni, derived with superconducting leads agree reasonably well with no-free-parameter calculations. The measurements and calculations of $\gamma_{Co/Cu}$ and $\gamma_{Co/Ni}$ agree fairly well, but the calculated values for $\gamma_{Fe/Cr}$ are a bit small. For $\gamma_{Py/Cu}$, agreement was found between superconducting leads and the same nanowire group as for $\gamma_{Co/Cu}$.

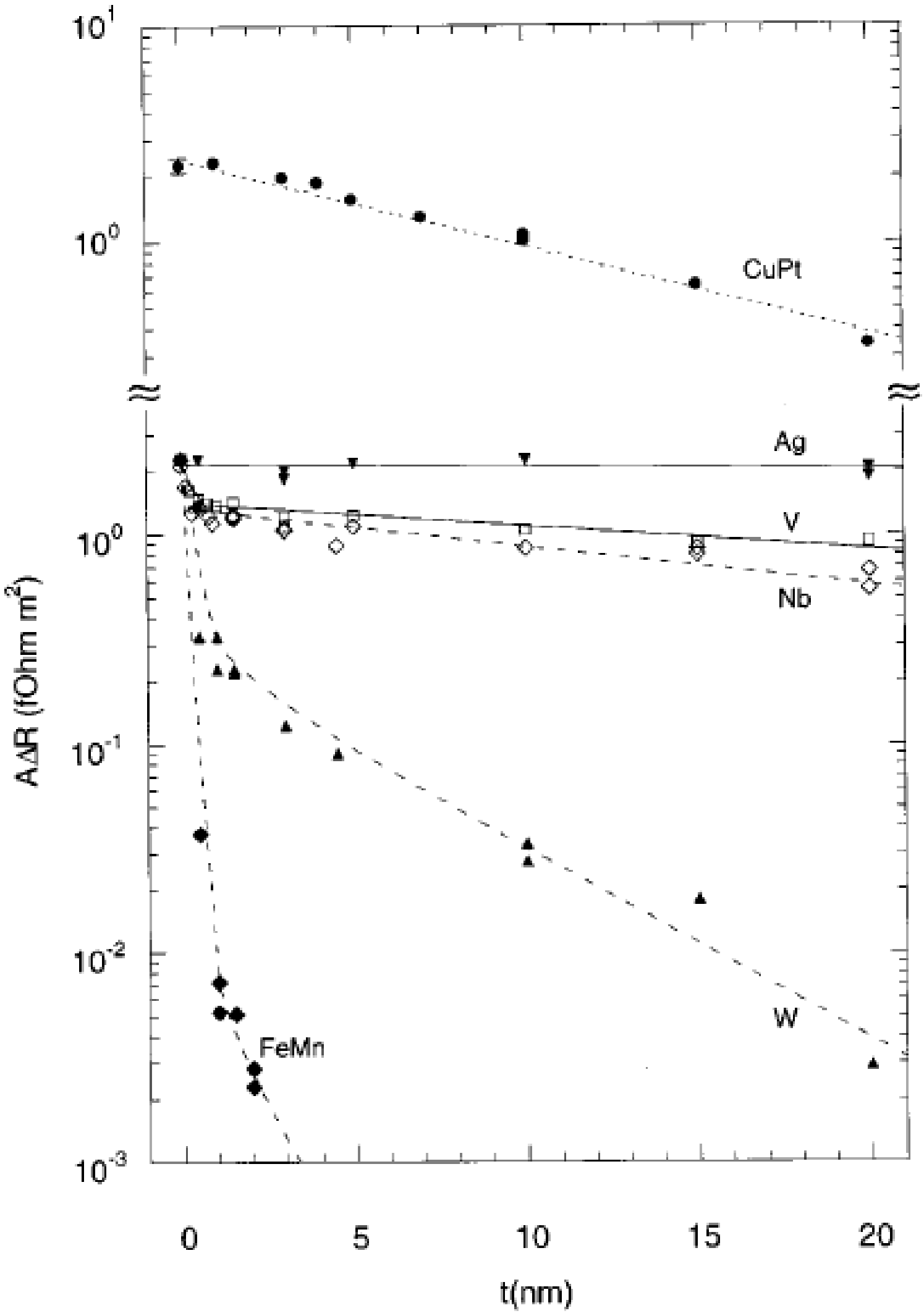

**Fig. 15**. AΔR vs t for sputtered Py-based EBSVs with inserts of the N-metals Ag, V, Nb, W, the alloy Cu(6%Pt), and the antiferromagnet FeMn. The slopes of the lines at 'large t' give the spin-diffusion-lengths: $l_{sf}^{CuPt}$ = 11 nm; $l_{sf}^{Ag}$ ≥ 40 nm; $l_{sf}^{V}$ ≥ 40 nm; $l_{sf}^{Nb}$ = 25 nm; $l_{sf}^{W}$ ≈ 4.8 nm; and for FeMn/Cu an effective interface spin-flipping parameter δ ≈ 2.5. The best values of $AR_{Cu/N}$ are (in fΩm$^2$): $AR_{Cu/Ag}$ = 0.044; $AR_{Cu/Nb}$ = 1.1; $AR_{Cu/V}$ = 1.15; $AR_{Cu/W}$ = 1.55. From W. Park et al. [20]

Two studies have looked for effects of changing interface physical roughness on CPP-MR, both with Fe/Cr. One reported an increase with increasing roughness [34]. The other reported an apparent slight decrease [98]. Table 5 shows that the value of $2AR^*_{Fe/Cr}$ derived in the second study agrees well with no-free-parameter calculations that take no account of physical roughness.

**(B) $2AR_{N1/N2}$.**

Values of the interface specific resistances of non-magnetic/non-magnetic (N1/N2) interfaces are determined using two different techniques. So far, all have been measured using superconducting cross-strips.

Method #1. The first method involves a multilayer with fixed total thickness $t_T$ that is divided into *n* equal thickness bilayers of N1 and N2 [24]. Since $t_T$ stays fixed, the total thicknesses of N1 and N2 also stay fixed at $t_T/2$, and increasing *n* simply increases the number of N1/N2 interfaces. To eliminate any proximity effect from the superconducting Nb on N1 and N2, the $[N1/N2]_n$ multilayer is sandwiched between 10 nm thick Co layers, giving: Nb/Co(10nm)/$[N1(t_T/2n)/N2(t_T/2n)]_n$/Co(10nm)/Nb. With values of $t_T$ = 360 nm or 540 nm, the two Co layers are so far apart that any magnetoresistance is negligible, as is checked by confirming no change in total AR ($AR_T$) with

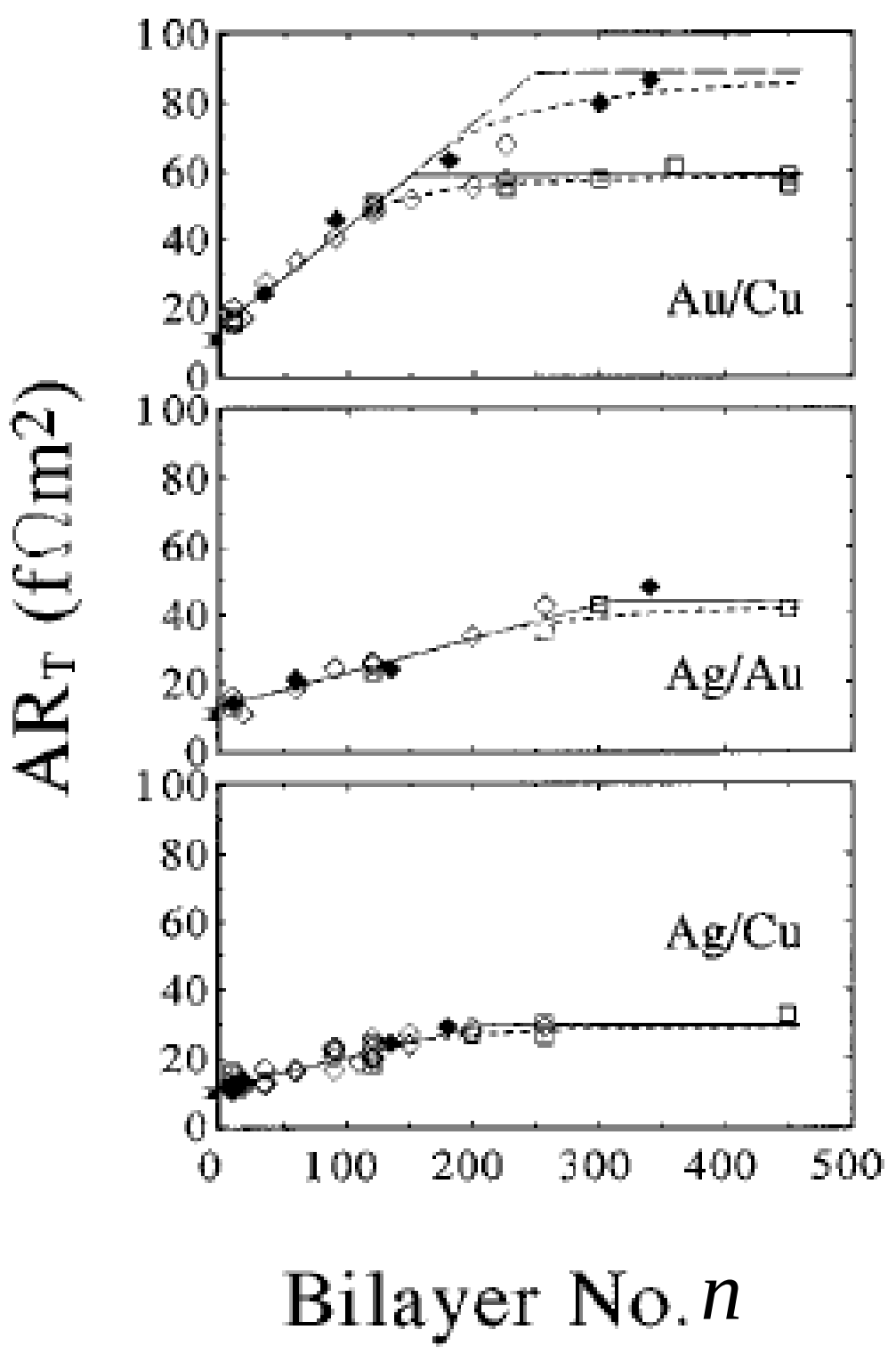

**Fig. 16**. $AR_T$ vs bilayer number $n$ for sputtered $t_T$ = 360 nm (open symbols) or 540 nm (filled symbols) thick $[N1/N2]_n$ multilayers of N1/N2 = Au/Cu, Ag/Au, or Ag/Cu with equal thicknesses of N1 and N2. Diamonds are for samples sputtered at standard rates; squares at half rates. The data should grow linearly until the layer thickness becomes less than the interface thickness and the slope of the linear region should give $2AR_{N1/N2}$. The best fit values are: $2AR_{Ag/Cu}$ = 0.09 f$\Omega$m$^2$; $2AR_{Ag/Au}$ = 0.1 f$\Omega$m$^2$; and $2AR_{Au/Cu}$ = 0.3 f$\Omega$m$^2$. The solid lines are for 'square' interface profiles. The dashed lines are for 'linear' interface profiles. From L.L. Henry et al. [24]

H for -500 ≤ H ≤ + 500 Oe. So long as the N1 and N2 layers are thicker than the thickness of the N1/N2 interface, the total sample specific resistance should be approximately:

$$AR_T = AR_{AP} = 2AR_{S/Co} + \rho^*_{Co}(20) + AR_{Co/N1} + AR_{Co/N2} + (\rho_{N1} + \rho_{N2})(t_T/2) - AR_{N1/N2} + 2nAR_{N1/N2}. \qquad (7)$$

A plot of $AR_T$ vs $n$ should then give a straight line up to where $n$ becomes large enough that the N1/N2 interfaces begin to overlap. For still larger $n$, the data should level off at a constant value corresponding to the AR expected for a 50%-50% alloy of N1 and N2.

Fig. 16 [24] shows such plots for Ag/Au, Ag/Cu, and Au/Cu multilayers, with the resulting values of $2AR_{N1/N2}$ given in the figure caption. Ref. [24] shows that the ordinate intercepts are consistent, to within mutual uncertainties, with the sums of the separately determined terms independent of $n$. The intersections of the extrapolated slopes and the 'constant' limits of large $n$ give estimates of the interface thicknesses, which correspond to ~ 0.6 nm for Ag/Au, ~ 0.9 nm for Ag/Cu, and ~ 1.2 nm for Au/Cu. These values are similar to, but larger than, the thicknesses estimated from x-ray measurements [24].

Method #2. The second method involves inserting a $[N1(3)/N2(3)]_n$ multilayer into the middle of a Py-based EBSV, giving : Nb(250)/Cu(10)/FeMn(8)/Py(24)/Cu(10)$[N1(3)/N2(3)]_n$Cu(10)/Py(24)/Cu(10)/Nb(250). Here, the Py layers are close enough to give a CPP-MR. The advantage of this technique is that measuring AΔR vs $n$ allows determination of $\delta_{N1/N2}$, the spin-relaxation parameter for an N1/N2 interface. The disadvantages for determining $2AR_{N1/N2}$ are that the 'constant background' is larger compared to the signal of interest, and that corrections must be made for spin-relaxation within the N1 and N2 layers. Fig. 17a [20] shows examples of AR vs $n$ for a series of Cu/N multilayers with N = Ag, V, Nb, and W. The values of AR in f$\Omega$m$^2$ (after corrections for the bulk contributions) are given in the caption to Fig. 15. A more extensive list of values of 2AR is given in [97].

Table 5 shows that, for lattice matched pairs, no-free-parameter calculations agree with measured values of $2AR_{N1/N2}$, whereas for non-lattice-matched pairs, the calculations and measurements disagree

**(C) $2AR_{S/F}$.**

Values of the interface specific resistances, $2AR_{S/F}$ of superconducting/ferromagnetic (S/F) interfaces, determined from the ordinate intercepts of data such as those in Fig. 3, are collected in Table 6 [17,25,51,74,81,83,92,98,104]. Intriguingly, all of the values for S = Nb determined in this way are very similar, ranging only from 4.8 to 7.5 f$\Omega$m$^2$ with uncertainties that almost all overlap with 6 ± 1 f$\Omega$m$^2$. These values are not sensitive to deposition of 5-10 nm of Ag or Cu between the Nb and the F layer [17,105], but are slightly sensitive to deposition of Au or Ru [105]. Satisfactory quantitative explanations for these results do not yet exist. Included in Table 6 are a larger $2AR_{S/F}$ for S = NbTi with a much larger residual resistivity than that of Nb, and three values determined by other techniques that we view as less reliable.

**(D) Spin-Relaxation at N1/N2, F/N, and F1/F2 interfaces: $\delta_{N1/N2}$, $\delta_{F/N}$, and $\delta_{F1/F2}$.**

Until recently, spin-relaxation at F/N interfaces has been neglected in CPP-MR analyses. For simple $[F/N]_n$ multilayers with large $n$, such neglect is appropriate, as VF analysis shows that adding $\delta_{F/N}$ ~ 0.2-0.3 typically changes $AR_{AP}$, $AR_P$ and AΔR by only a few percent. In contrast, adding such values of $\delta_{F/N}$ to EBSVs tends to have a larger effect, as we'll discuss below A technique for measuring $\delta_{N1/N2}$ was developed and applied back in 2000 [20]. In contrast, a technique to specifically measure $\delta_{F/N}$ or $\delta_{F1/F2}$ was developed only recently [21], and Table 7 [21,101,104,106] gives a collection of such values.

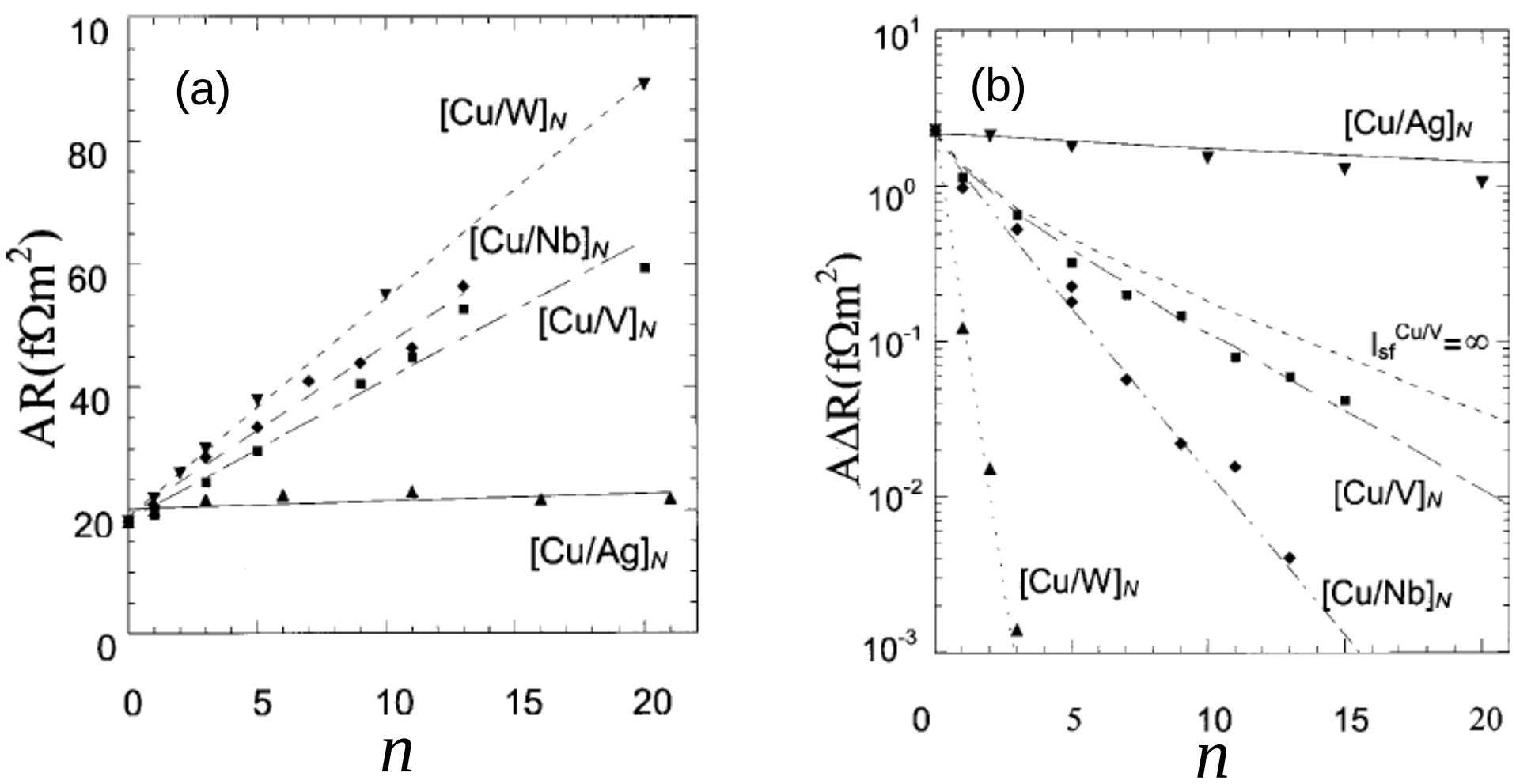

**Fig. 17**. (a) $AR_{AP}$ vs bilayer number *n* for $[N1/N2]_n$ multilayers sandwiched between Py-layers in a Py-based exchange-biased spin-valve. The slopes of the lines give $2AR_{N1/N2}$, after correction for the resistivities of N1 and N2. (b) AΔR vs *n* for the same samples. The slopes of these lines give $\delta_{N1/N2}$ after correcting for $l_{sf}$ in the bulk metals. The best values of the parameters derived from Fig. 17b are: $\delta_{Cu/Ag} = 0$; $\delta_{Cu/V} = 0.07$; $\delta_{Cu/Nb} = 0.19$; $\delta_{Cu/W} = 0.96$. From W. Park et al. [20].

**Table 5. $2AR_{N1/N2}$ or $2AR^*_{F/N}$ , and $\gamma_{F/N}$, for lattice matched and some mismatched pairs at 4.2K.**
Values are rounded to significant figures. Units for 2AR are fΩm$^2$ . Orientations = (111) for fcc and (011) for bcc. Calculations are for perfect (flat) interfaces with no mixing, or for 2ML of a 50%-50% alloy. Δa/a(%) is the % difference in lattice parameters for the two metals.

| | | | | | | | |
|---|---|---|---|---|---|---|---|
| | | | Matched Pairs | | | | |
| Metals | Δa/a(%) | 2AR(exp) | 2AR(perf.) | 2AR(50-50) | γ(exp). | γ(perf) | γ(50-50) |
| Ag/Au | 0.2 | 0.1 [24] | 0.09 [60,62] | 0.13[60,62] | | | |
| Co/Cu* | 1.8 | 1.0 [4] | 0.9 [60,62] | 1.1 [60,62] | 0.8 [4] | 0.6 [60,61] | 0.6 [60] |
| Fe/Cr* | 0.4 | 1.6 [98] | 1.7 [64] | 1.5 [64] | -0.7[98] | -0.5[60,61] | -0.3 [60] |
| Pd/Pt | 0.8 | 0.3 [63] | 0.4 [64] | 0.4 [64] | | | |
| Pd/Ir | 1.3 | 1.0 [64] | 1.1 [64] | 1.1 [64] | | | |
| Co/Ni | 0.6 | 0.5 [104] | 0.4 [104] | 0.4 [104] | 0.94[104] | 0.96[104] | 0.96[104] |
| | | | Mismatched Pairs | | | | |
| Ag/Cu | 12 | 0.09[24] | 0.45 [103] | 0.6 [103] | | | |
| Au/Cu | 12 | 0.3 [24] | 0.45 [103] | 0.7 [103] | | | |
| Pd/Cu | 7 | 0.9 [102,103] | 1.5 [103] | 1.6 [103] | | | |

**(D1) $\delta_{N1/N2}$.**

As noted in section VIII.B, method #2 for finding $2AR_{N1/N2}$, involving insertion of an $[N1/N2]_n$ multilayer into the middle of a Py/Cu/Py EBSV, also allows determination of $\delta_{N1/N2}$ by measuring how AΔR varies with *n*. Fig. 17b [20] shows examples of such data for Cu/N pairs. The slope of the exponential decay with increasing *n* gives the sum of three contributions: slope = $-(t_{N1}/l_{sf}^{N1} + t_{N2}/l_{sf}^{N2} + 2\delta_{N1/N2})$, where the factor of 2 is for two N1/N2 interfaces for each N1/N2 pair. The figure caption gives the values of $\delta_{N1/N2}$ for the Cu/N pairs shown. Ref. [72] lists the pairs published so far.

**(D2) $\delta_{F/N}$ and $\delta_{F1/F2}$.**

**Table 6. 2AR$_{F/S}$ and superconductor resistivity, $\rho_S$ above $T_c$.** We view the values in italics as less reliable.

| F-metal | S-metal | 2AR$_{S/F}$ ($10^{-15}$ $\Omega m^2$) | $\rho_S$ ($10^{-8}$ $\Omega$m) |
|---|---|---|---|
| **Co** | **Nb** | **6.1 ± 0.3 [17]** | **~ 6** |
| **Co** | **Nb** | **6 ± 1 [25]** | **~ 6** |
| **Ni** | **Nb** | **4.8 ± 0.6 [17]** | **~ 6** |
| **Ni** | **Nb** | **5 ± 1 [104]** | **~ 6** |
| **Fe** | **Nb** | **7.2 ± 0.5 [92]** | |
| **Fe** | **Nb** | **6 ± 1 [98]** | **~ 6** |
| **$Co_{91}Fe_9$** | **Nb** | **7 ± 1 [83]** | **~ 6** |
| **Py = $Ni_{84}Fe_{16}$** | **Nb** | **6 ± 1 [51,74]** | **~ 6** |
| **Py' = $Ni_{66}Fe_{13}Co_{21}$** | **Nb** | **7.5 ± 1 [81]** | **~ 6** |
| **Co** | **NbTi** | **12.4 ± 0.7 [17].** | **~ 57** |
| **FeMn** | **Nb** | **2.0 ± 1.2 [74].** | **~ 6** |
| ***NiCr*** | ***Nb*** | ***15 ±4 [17].*** | ***~ 6*** |
| ***Cu*** | ***Pb(Bi)*** | ***3.5 [17].*** | ***~ 5*** |
| ***W*** | ***In(Pb)*** | ***7 [17].*** | ***~ 2*** |

**Table 7. $\delta_{F/N}$ and $\delta_{F1/F2}$.**

| **Metal Pair** | **Δa/a (%).** | **$\delta_{F/N}$ or $\delta_{F1/F2}$.** |
|---|---|---|
| Co/Cu | 1.8 | $0.33^{+0.03}_{-0.08}$ [21] |
| Co/Ni | 0.05 | 0.35 ± 0.05 [104] |
| $Co_{90}Fe_{10}$/Cu | 1.8 | 0.19 ± 0.04 [101] |
| Co/Ru | 7 | $0.34^{+0.04}_{-0.02}$ [106] |

A non-magnetic $[N1/N2]_n$ multilayer inserted into the middle of an EBSV does not perturb the magnetic structure and behavior of the EBSV, but just scatters (and flips the moments of) the electrons transiting from the pinned F-layer to the free one. In contrast, inserting a magnetic $[F/N]_nF$ or $[F1/F2]_nF1$ multilayer into the middle of an EBSV changes the magnetic structure. A technique that changes this structure in a controlled way to allow derivation of $\delta_{F/N}$ or $\delta_{F1/F2}$ is described in ref. [21]. The technique requires producing an $[F/N]_nF$ or $[F1/F2]_nF1$ multilayer with the F-layers ferromagnetically coupled so that the multilayer reverses as a single unit. This multilayer is then embedded in the middle of a symmetric, Py-based double exchange-biased spin-valve (DEBSV), giving something close to two EBSVs in series. The symmetric DEBSV gives approximately twice the signal of a single EBSV. For the $[F/N]_nF$ multilayer, the thickness of N is chosen to give ferromagnetic coupling. For the $[F1/F2]_n$ multilayer, exchange coupling should be ferromagnetic. In the studies so far, F and N, or F1 and F2, and their thicknesses, have been chosen so that spin-relaxation (spin-diffusion) in the bulk is weak enough to allow isolation of the contribution of spin-relaxation from the interfaces. $\delta_{F/N}$ or $\delta_{F1/F2}$ is found by measuring AΔR vs $n$, as illustrated in Fig. 18 [21] for: $[Co(3)/Cu(1.3 \text{ or } 1.5)]_n/Co(3)$. The dashed curve shows the behavior of AΔR expected for $\delta_{Co/Cu} = 0$. The solid curve is a best fit to the data with $\delta_{Co/Cu} = 0.33$. Published values of $\delta_{F/N}$ and $\delta_{F1/F2}$ are given in Table. 7. The main caveat is that most rely on the assumption (based on experiment [23,75]) of a long spin-diffusion length in Co ($l_{sf}^{Co} \sim$ 60 nm at 4.2K).

As yet, understanding of δ is minimal. We don't know if $\delta_{N1/N2}$, $\delta_{F/N}$,, or $\delta_{F1/F2} \neq 0$ can occur for perfect interfaces due to differences in spin-orbit parameters of the two metals, or if they require alloyed interfaces. For $\delta_{F/N}$ and $\delta_{F1/F2}$, magnetic disorder at the interface may be an additional contributor [107].

**IX. Work toward CPP-MR Devices.**

Present Tunneling Magnetoresistance (TMR) (~ 500 Gb/in$^2$) read heads have resistances ~ 500 Ω and TMR ~ 50% -100% with values of AR extending to below 1 Ω(μm)$^2$ and smallest dimensions ~ 50 nm. Typically, TMR devices with smaller values of AR have reduced TMR. Metallic CPP-MR devices offer the potentials of lower AR and lower Johnson noise [108], but not yet large and reproducible enough CPP-MRs in devices with all the required characteristics to supplant TMR devices in commercial read heads. Fig. 19 [109] compares calculated Head-Amp Signal to Noise Ratio (SNR) vs sensor width of a TMR head with assumed MR = 50% and AR = 1 Ω(μm)$^2$, a "Current-Screened' = Current-Confined-Path (CCP) multilayer with assumed MR = 20% and AR = 0.25 Ω(μm)$^2$, and an all-metal CPP-MR multilayer with assumed AR = 0.04 Ω(μm)$^2$ and MR = 10%. In that case, an all-metal multilayer would be preferred for sensor widths below ~ 35 nm.

**(A) F-layer lamination.**

Table 3 shows that $\gamma_{Co/Cu} > \beta_{Co}$, and $2AR^*_{Co/Cu} > \rho^*_{Co}t_{Co}$ for $t_{Co} \leq 12$ nm. Thus, laminating 3 nm or 6 nm thick Co layers into $n$ ferromagnetically coupled thinner Co/Cu bilayers, by inserting $n$ thin (0.5 nm) Cu layers into the Co, should increase both AΔR and $AR_P$, without greatly increasing the multilayer thickness. Eid et al. [110] and Delille et al. [111] both found such lamination to increase both AΔR (by as much as 100%) and $AR_P$, but by increasingly less than predicted as $n$ grew. Eid et al. [110] tentatively attributed the weakened increase with $n$ to spin-relaxation at the Co/Cu interface (see section VIII.D). Room temperature values of AΔR ~ 0.005 $\Omega(\mu m)^2$ with AR ~ 0.13 $\Omega(\mu m)^2$ were reported upon inserting half-ML Cu 'layers' into $Co_{50}Fe_{50}$-layers, although the results were not attributed to lamination [112].

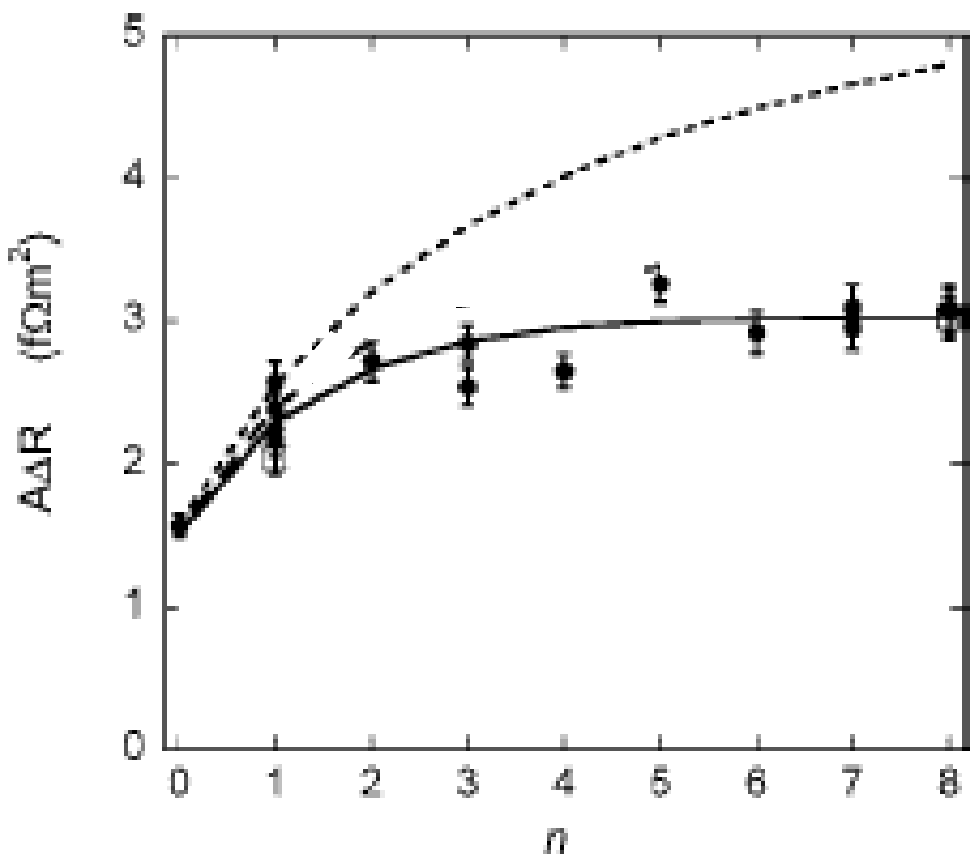

**Fig. 18.**, AΔR vs *n* for Py-based double exchange-biased spin-valves (DEBSVs) with ferromagnetically coupled inserts of $[Co(3)/Cu(t_{Cu})]_nCo(3)$. Filled circles are $t_{Cu}$ = 1.3 nm and open squares are $t_{Cu}$ = 1.5 nm. The solid curve is the best fit with the parameters given in the original paper and the value of δ = 0.33 (see Table 7). The dashed curve is calculated with the same parameters but assuming δ = 0. After B. Dassonneville et al. [21].

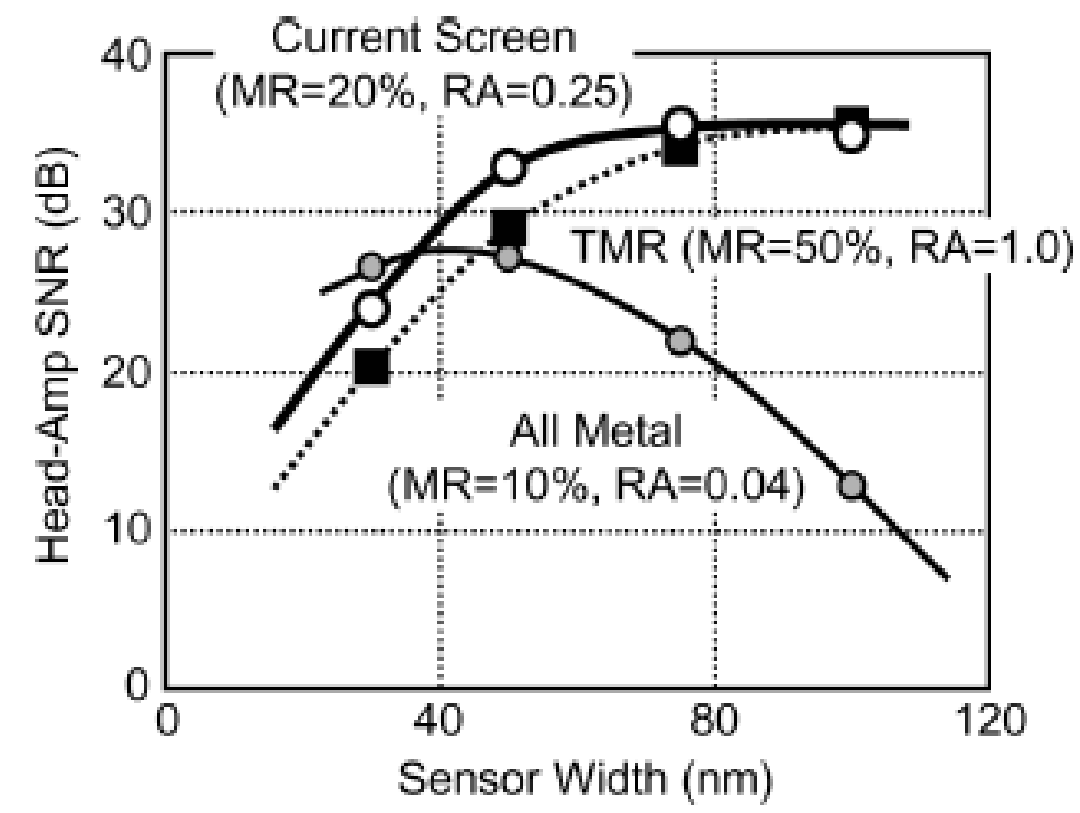

Fig. 19. Calculated head-amp signal to noise ratio (SNR) for TMR, Current Screen = NOL, and All Metal Read Heads with the MR% and with AR in units of $\Omega(\mu m)^2 = 10^3$ $f\Omega m^2$. From K. Nakamoto et al. [109].

**(B) Current Confined Paths (CCP) via Nano-oxide layers (NOL).**

As oxidizing the surfaces of CIP multilayers to get specular reflection increases the CIP-MR [113], Nagasaka et al. [114] reasoned that oxidizing within the layers of a CPP-MR sample might give specular reflection and increase the CPP-MR. Indeed, they found such oxidation to increase the CPP-MR, but. Ffurther work showed that thethe source was not speciular reflection. Rather, the best results came from inserting a thin layer of $Al_2O_3$ within the Cu spacer to give an insulating nano-oxide layer (NOL) with small conducting channels that give Current-Confined-Paths (CCP). Such CCPs have been reported to give both larger $AR_P$ and larger AΔR [109,115,116]. Fig. 19 [109] compares calculated values of signal-to-noise ratios (SNR) for such a CCP device, with TMR and all metallic CPP-MR devices with plausible assumed values of AR and MR%. Hydrogen ion treatment (HIT) increased the CPP-MR of a CCP-based multilayer to about 25% for AR ~ 500 $f\Omega m^2$ [117]. To be competitive for devices, both the size and distribution of channels would have to be well controlled in nm scale pillars, control not yet demonstrated.

**(C) F-alloys or Compounds To Give Large Room Temperature CPP-MR.**

Fig. 19 shows that an all-metal CPP-MR device with low AR ~ 40 $f\Omega m^2$ could be favored for ultra-high density recording (track dimension ≤ 35 nm) if a CPP-MR ≥ 10% can be reproducibly achieved in a multilayer subject to such a size constraint. Great effort has been expended to try to find multilayers with F-metals, alloys, or compounds to meet this need. To do so requires F- and N-layers with $\beta_F$ and/or $\gamma_F$ closer to 1. In principle, a perfect half-metallic F-layer of compounds such as Heusler alloys, where only electrons of one moment direction can propagate, should give $\beta_F = 1$. So far, no-one has yet fabricated F/N CPP-MR multilayers with $\beta_F$ or $\gamma_F$ very near 1. An early Heusler alloy study gave only a modest CPP-MR [118]. Later studies of Heusler alloys and other compounds with various spacer metals, have given values of AΔR, CPP-MR, and AR close enough to the requirements that CPP-MR looks to be a serious competitor for next-generation read heads. Included are reports of room temperature values of AΔR from 5 – 10 $f\Omega m^2$, CPP-MRs from 11% to 42%, and ARs from 20 – 200 $f\Omega m^2$, [119-13229]. Still to be fully addressed are spin-torque induced excitations and noise in such multilayers [12930-132].

**X. Magnetothermoelectricity and Thermal Conductance.**

Early studies [37,48,49] showed that Magnetothermoelectric Power data for F/N multilayers in the CPP geometry behaved similarly to CPP-MR data, with modestly larger fractional changes with H. Data on more complex phenomena, such as a spin-dependent Peltier effect [37], including when both heat and charge currents are flowing [37], as well as effects involving non-collinear spin-orientations [1330], have been reported. Fig. 20 [37] compares CPP-MR, CPP-Magnetothermopower (MTP), and an ac CPP Magneto-thermalgalvanic Voltage (MTGV) signal detected with both an ac applied temperature difference and a constant 200 μA current through the sample. The larger MTGV signal is attributed to the Peltier effect, with spin-dependence of the Peltier coefficient [37]. Most of these phenomena are still under theoretical study [1341,1352]; see articles in the Spin Caloritronics issue of Solid State Comm. [1363].

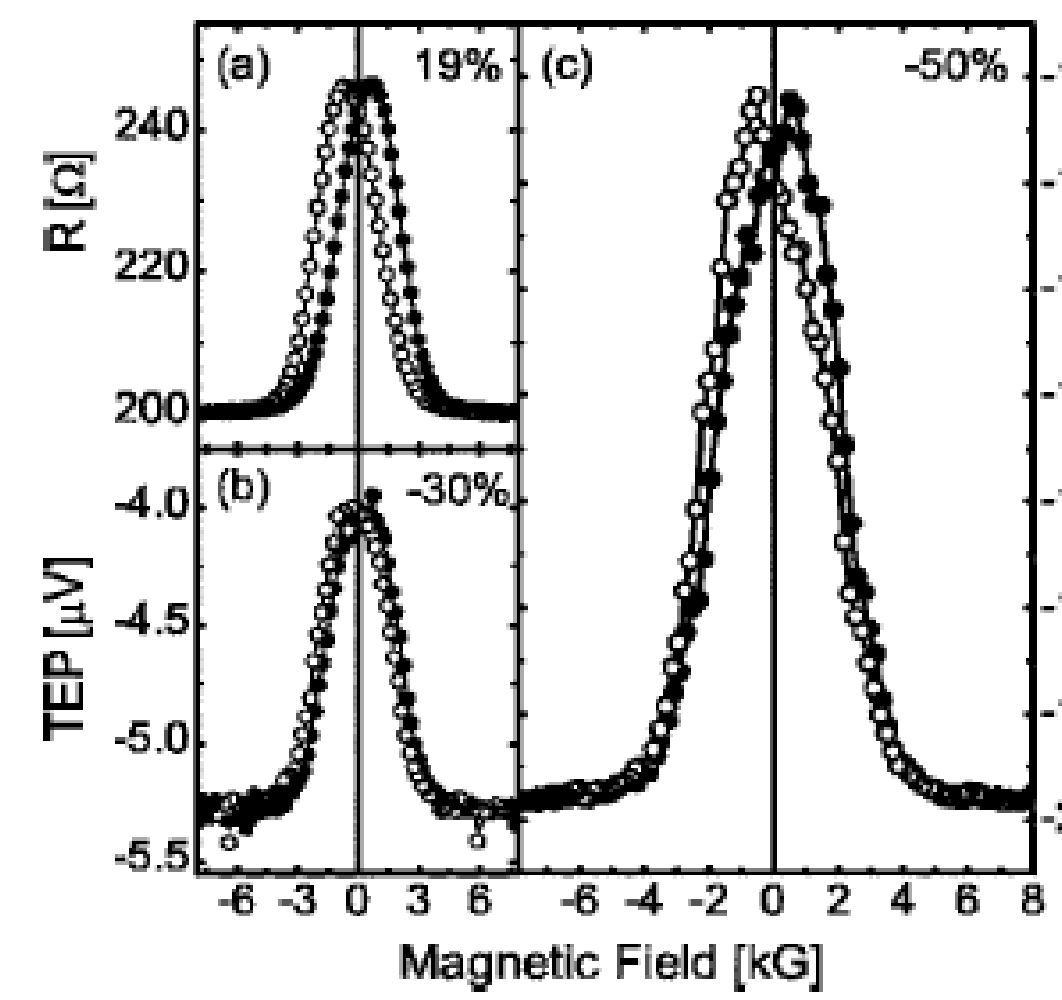

Fig. 20. (a) CPP-MR; (b) CPP-MTEP; (c) CPP-MTGV curves at 15K. The a.c. MTGV data were measured under a dc current of -200 μA. Full (open) symbols indicate field sweep up (down). From L. Gravier et al. [37].

An additional quantitative CPP-MR result: the thermal conductances of sputtered Pd/Ir interfaces at temperatures from 78-295K were found [1374] to agree with the Wiedemann-Franz law to within 10%, assuming a temperature independent interface specific resistance for the same samples at 4.2K [64].

**XI. Summary.**

We have shown that most (perhaps all) CPP-MR data can be interpreted consistently using Valet-Fert (VF) theory and its parameters, based upon assuming dominance of diffuse scattering. Especially important results include the following, mostly at T = 4.2K. The few studies of temperature dependences suggest that the temperature variations of $\beta_F$, $\gamma_{F/N}$, and $2AR^*_{F/N}$, are usually modest ( ≤ 10-20%)[47,95,125]. (1) Data for a variety of different multilayers are consistent with the forms predicted by VF theory and, when spin-relaxation is very weak, with its two-current series-resistor (2CSR) model limit. (2) The most carefully derived CPP-MR values of the bulk asymmetry parameter, $\beta_F$, for F-based alloys are consistent with those derived earlier from Deviations from Matthiessen's Rule studies. (3) CPP-MR spin-diffusion lengths in well characterized N-alloys agree with those derived from independent electron spin-resonance measurements of spin-orbit coupling. (4) Spin-diffusion lengths in the F-alloy Py measured independently by different groups mostly agree, and the best values agree with one calculated with no adjustment. (5) Most spin-diffusion lengths in F-alloys measured so far fall close to a single line characterizing a linear relation between the spin-diffusion lengths and the mean-free-paths (inverses of residual resistivities). (6) Spin-relaxation at the interfaces of sputtered antiferromagnets FeMn and IrMn with Cu is strong, and the spin-diffusion lengths in sputtered bulk FeMn and IrMn are probably short. (7) The values of interface specific resistances, 2AR or $2AR^*$, for lattice matched metal pairs agree well with no-free-parameter calculations and are often not sensitive to interface intermixing. (8) The few published calculations of the interface asymmetry parameter, $\gamma_{F/N}$ or $\gamma_{F1/F2}$, are roughly consistent with measured values. (9) Non-zero values have been reported for interface spin-relaxation parameters, $\delta_{N1/N2}$, $\delta_{F/N}$, and $\delta_{F1/F2}$. (10) Significant magnetothermoelectric effects have been seen in the CPP geometry. (11) Recent progress in enhancing CPP-MR with new ferromagnetic alloys or compounds, and by techniques such as current-confined paths (CCP), gives promise that metallic CPP-MR might supplant tunneling MR (TMR) in commercial read heads as bit density increases.

Phenomena still to be understood include non-zero values of δ at various interfaces, the observed values of $2AR_{S/F}$, and magnetothermoelectric effects. Also not yet clear are effects of physical roughness (as distinct from intermixing) upon interface properties. Further work on materials is needed to make CPP-MR still larger. Finally, it would be interesting to make samples free enough from disorder to give clear evidence for ballistic transport in CPP-MR.

**Supplementary note #1: Contact resistances and non-uniform current flow in micro- and nanopillars.**

Measurements of $AR_{AP}$, $AR_P$, $A\Delta R$, and CPP-MR on micro- or nanopillars are subject to two types of errors: (1) Contact resistance, $R_c$, and (2) Non-uniform current flow through the pillar. Here we explain the sources of these errors and give closed form approximations sufficient to estimate the sizes of both effects.

Consider the usual geometry of a cylindrical pillar of small radius r sandwiched between two, much wider (W >> r) and longer (L >> r) thin contact films. The pillar has area $A = \pi r^2$, length $\ell$, and average resistivity $\rho_A$, giving specific resistance $AR_o = \rho_A \ell$. ($AR_o$ can be $AR_{AP}$ or $AR_P$). The top and bottom contact films each have the same resistivity $\rho_t = \rho_b$ and the same thickness $t_t = t_b$, thus giving a total sheet resistance $R_s = 2\rho_t/t_t$. The current and voltage contacts are located L/2 >> r away from the pillar. (Note: an alternative geometry is described in [35,139]).

(1) Current crowding. For the chosen geometry, ref. [138] gives an approximate analytical expression $R_c \approx 0.1R_s \ln(W/r)$. With values of W ≥ 100 μm and r ≤ 10 μm, $R_c$ will be comparable to $R_s$. Since this $R_c$ is the same for $AR_{AP}$ and $AR_P$, it will not affect $A\Delta R$, but will reduce the CPP-MR. Because the 'ln' term varies weakly with (1/r) $\propto (1/\sqrt{A})$, for a small enough range of 1/A, a plot of the measured $AR_m$ vs 1/A might approximate a straight line with slope $AR_P$ or $AR_{AP}$ [112].

(2) Non-uniform current. Qualitatively, in the chosen geometry, current flows into the top of the pillar uniformly across its circumference, and then flows from the top to the bottom of the pillar uniformly if $R_s$ is sufficiently small, but only in an annulus of approximate thickness λ (defined below) if $R_s$ is large. In the latter case, the current density just inside the radius r increases as λ decreases, and both $AR_m$ and $A\Delta R_m$ increase as we now describe. Ref. [35] used a two-dimensional analysis to derive an approximate expression for the measured $AR_m = C(x/2)I_o(x)/I_1(x)$, where $I_o$ and $I_1$ are modified Bessel functions of zeroth and 1st order. x, which determines the non-uniformity of the current through the pillar, is given by x = r/λ, where $\lambda = \sqrt{(C/R_s)}$ and $C = [(1/2)(2\rho_t t_t) + \rho_A \ell\,] = \rho_t t_t + AR_o$. As noted above, the length λ determines the 'thickness' of the current flow down through the pillar. We give three examples of how $AR_m$ varies with x. As $x \to 0$ ($\lambda \to \infty$), $I_o(x) \to 1$, $I_1(x) \to x/2$, and $AR_m \to C$. As $x \to \infty$ ($\lambda \to 0$), $I_o(x)/I_1(x) \to 1$ and $AR_m \to Cx/2$ which grows as $\sqrt{A}$. For x = 1, $I_o(1)/I_1(1) = 2.24$, and $AR_m = 1.12C$. Note that C is larger than the desired $AR_o$ by $\rho_t t_t = R_s(t_t)^2$, and thus $AR_m$ must be corrected if $R_s(t_t)^2$ is not << $AR_o$.

To conclude, values of $AR_{AP}$, $AR_P$, $A\Delta R$, and CPP-MR directly measured on pillars are reliable only if the total sheet resistance of the contact films is much less than the resistances $R_{AP}$ and $R_P$ of the pillar. In general this means that the smaller the pillar diameter, and the greater the thickness and lower the resistivity of the contact films, the better. If either one or both of the problems described above occur, corrections are required to obtain reliable values of $AR_{AP}$, $AR_P$, $A\Delta R$, and CPP-MR.

**Suppl. Note #2. Spin-Diffusion and Related Lengths Determined by Different Techniques.**

Here we explain how those values of $l_s$ in Table 1 specified as calculated from conduction electron spin-resonance (CESR) cross-sections, $\sigma_{sf}$ [71], were obtained, correct two errors in ref. [72], and specify how to properly compare spin-diffusion lengths in non-magnetic metals measured by transport properties such as CPP-MR with spin-orbit lengths measured by weak localization. Following Zutic et al. [140] we define four relaxation times, the momentum relaxation time, τ, the spin-orbit time $\tau_{so}$, the spin-flip time, $\tau_{sf}$, and the spin relaxation time, $\tau_s$. The three times involving spin-flips are related to each other by $\tau_{sf} = 2\tau_s$ [11,140] and (for an isotropic system) $\tau_{so} = (4/3)\tau_s$[140,141]. Each relaxation time is related to a 'mean-free-path' by $\lambda = v_F\tau$, where $v_F$ is the Fermi velocity.

The general equation relating $l$ to τ is $l_x = \sqrt{D\tau_x}$, where $D = (1/3)(v_F\lambda)$ and x specifies the same quantity for $l$ and τ. Thus, for 's', we have $l_s = \sqrt{D\tau_s} = \sqrt{(1/3)(\lambda\lambda_s)}$ , (A1.1) for 'so', we have $l_{so} = \sqrt{D\tau_{so}} = \sqrt{(1/3)(\lambda\lambda_{so})} = \sqrt{(4/9)(\lambda\lambda_s)} = (2/\sqrt{3})l_s$, (A1.2) and for 'sf', we have $l_{sf} = l_s(\mathrm{CESR}) = \sqrt{D\tau_s} = \sqrt{(1/3)(\lambda\lambda_s)} = \sqrt{(1/6)(\lambda\lambda_{sf})}$ [19] (A1.3).

Eq. A1.3 says that equation (4) of ref. [72] describing how to calculate $\lambda_{sf}^N$ for a dilute alloy of host metal N with impurity concentration $c$ and number of host atoms per unit volume, $n$, from CESR measurements of $\sigma_{sf}$ should be simplified to $\lambda_{sf}^N = [1/nc\sigma_{sf}] = [1/nc\sigma_{\mathrm{CESR}}]$. Eq. A1.3 was used for the calculations in both Table 1 of the present chapter and Table 1 of ref. [72]; thus the comparisons made in both of those tables are correct as given.

In contrast, Eq. A.1.2, says that, as demonstrated by Niimi et al. [141], those values of $l_{so}$ measured by weak-localization and listed in Table 2 of [72] should be corrected by $\sqrt{3}/2$ before being compared with the listed values of $l_s$ as measured by transport.

**Acknowledgments.** Some of the research described in this review received funding from the US-NSF Division of Materials Research (DMR), from Seagate Inc., and from the Korean Institute of Science and Technology (KIST).

The author is pleased to acknowledge helpful comments and suggestions from: N.O. Birge, J.R. Childress, J. Fabian, A. Fert, Y. Otani, W.P. Pratt Jr., and I. Zutic.

## CPP-MR References